\RequirePackage{lineno}
\documentclass[aps,prl,final,twocolumn,letterpaper, superscriptaddress]{revtex4-2}

\usepackage{graphicx}   
\usepackage{import}                         
\usepackage{epstopdf}
\usepackage{amsmath} 
\usepackage{float} 
\usepackage{bm}
\usepackage{amssymb}
\usepackage{quotes}
\usepackage{indentfirst}
\usepackage{color}
\usepackage[utf8]{inputenc}
\usepackage{transparent}
\usepackage{dcolumn}
\usepackage{braket}
\usepackage{multirow}
\usepackage{cancel} 
\usepackage{mdframed}
\usepackage{color}
\usepackage{bm}
\usepackage{dsfont}

\usepackage{slashed}
\usepackage{graphicx}
\usepackage{amsmath}
\usepackage{siunitx}
\usepackage{hyperref}
\usepackage{cleveref}
\crefname{equation}{Eq.}{Eqs.}
\crefname{figure}{Fig.}{Figs.}
\usepackage{soul, color} 
\soulregister\ref{7}  
\soulregister\cite{7} 
\renewcommand{\st}[1]{}

\usepackage{xr}
\usepackage{subfiles}

\makeatletter
\newcommand*{\addFileDependency}[1]{
  \typeout{(#1)}
  \@addtofilelist{#1}
  \IfFileExists{#1}{}{\typeout{No file #1.}}
}
\makeatother

\usepackage{textcomp} 
\usepackage{xifthen}
\usepackage{xcolor}
\usepackage{etoolbox}
\newboolean{togglechanges} 

\setboolean{togglechanges}{false}

\newcommand{\comment}[1]{\ifbool{togglechanges}
    {#1}  
    {\textcolor{blue}{#1}}}

\usepackage{bibentry}

\usepackage{graphicx}
\usepackage{dcolumn}
\usepackage{bm}

\usepackage{textcomp} 

\makeatletter
\renewcommand{\fnum@figure}{\textbf{Fig.~\thefigure}}
\makeatother

\makeatletter

\renewcommand{\makeLineNumber}{%
  \ifnum\pagegrid@cur=1\relax
    \makeLineNumberLeft
  \else
    \makeLineNumberRight
  \fi
}

\makeatother

\begin{document}
\linenumbers
\rmfamily

\title{Nanophotonic control of spatial information in scintillation detectors}

\author{Joshua~Chen}
\email{chenjosh@mit.edu}
\affiliation{Research Laboratory of Electronics, MIT, Cambridge, MA 02139, USA}

\author{Simo~Pajovic}
\affiliation{Department of Mechanical Engineering, MIT, Cambridge, MA 02139, USA}

\author{Seou~Choi}
\affiliation{Research Laboratory of Electronics, MIT, Cambridge, MA 02139, USA}

\author{Sachin~Vaidya}
\affiliation{Research Laboratory of Electronics, MIT, Cambridge, MA 02139, USA}
\affiliation{Department of Physics, MIT, Cambridge, MA 02139, USA}

\author{William~Michaels}
\affiliation{Research Laboratory of Electronics, MIT, Cambridge, MA 02139, USA}

\author{Louis~Martin-Monier}
\affiliation{Department of Materials Science and Engineering, MIT, Cambridge, MA 02139, USA}

\author{Christina~M.~Spägele}
\affiliation{Harvard John A. Paulson SEAS, Harvard University, Cambridge, MA 02138, USA}

\author{Steven~E.~Kooi}
\affiliation{Institute for Soldier Nanotechnologies, Massachusetts Institute of Technology, Cambridge, MA 02139, USA}

\author{Juejun~Hu}
\affiliation{Department of Materials Science and Engineering, MIT, Cambridge, MA 02139, USA}

\author{Rajiv~Gupta}
\affiliation{Neuroradiology Division, Department of Radiology, Massachusetts General Hospital, Harvard Medical School, Boston, MA 02114, USA}

\author{Charles~Roques-Carmes}
\affiliation{Institute of Science and Technology Austria (ISTA), Klosterneuburg 3400, Austria}

\author{Marin~Soljačić}
\affiliation{Research Laboratory of Electronics, MIT, Cambridge, MA 02139, USA}
\affiliation{Department of Physics, MIT, Cambridge, MA 02139, USA}



\clearpage 




\begin{abstract}
\begin{linenumbers}
X-rays enable non-invasive imaging across medicine, security, materials science, and beyond, yet modern systems remain constrained by the need to resolve finer structures at lower radiation dose~\cite{smith2025projected}. Scintillators are the dominant materials for detecting X-rays, but face a longstanding compromise: thick scintillators absorb X-rays effectively, whereas optical photons generated throughout their volume spread before detection, degrading spatial information~\cite{pelc2014recent}. Existing scintillator architectures largely try to preserve resolution by physically confining light using pixels~\cite{pelc2014recent,baffour2022ultra}, columnar crystals~\cite{miller2005recent}, or microstructured channels~\cite{ohashi2013submicron,hormozan2016high}. Here, we show that high-resolution detection does not require the volumetric confinement of scintillation light. A metalens integrated directly with a bulk scintillator uses nanophotonic wavefront control~\cite{Chen2026WavefrontEngineering} to preferentially transfer high-spatial-frequency information from volumetrically generated scintillation light to the detector, while retaining the X-ray absorption of a thick scintillator. We experimentally recover fine spatial detail in X-ray images of inorganic and biological specimens. In a detector geometry relevant to computed tomography (CT), the experimentally validated model predicts a fivefold reduction in required X-ray dose and a 25-fold increase in resolution bandwidth relative to a state-of-the-art pixelated scintillator. These results establish wavefront engineering as a route to separating efficient X-ray absorption from optical image formation, with the potential for substantially higher-resolution, lower-dose CT.
\end{linenumbers}
\end{abstract}

\maketitle

\section{Introduction}

X-ray imaging enables non-invasive inspection across medicine, security, industry, and materials science. In medicine, an outstanding challenge is to make X-ray detectors that resolve smaller structures at lower radiation dose. This is particularly important in computed tomography (CT), which reconstructs three-dimensional anatomy from X-ray projections. In the United States, an estimated 93 million CT examinations were performed in 2023, and the associated radiation exposure was projected to cause nearly 103,000 future cancers~\cite{smith2025projected}. At the same time, clinically important structures can lie below current CT resolution, including coronary plaque microcalcifications smaller than 60~µm~\cite{vancheri2019coronary}, small airways below 0.5~mm~\cite{koo2018small}, and suspicious breast microcalcifications tens to hundreds of micrometres across~\cite{wang2014non}. Nearly all current clinically-deployed CT scanners use indirect detectors, in which a scintillator converts X-rays into optical photons that are subsequently recorded by photodetectors. 


Detecting the fine anatomical features described above requires an X-ray detector to preserve the high spatial frequencies of the incident projection (Fig.~1a). In an indirect detector, this information must survive conversion into scintillation light and propagation through the scintillator~\cite{cunningham1994spatial, shultzman2023enhanced}. Efficient absorption of the X-ray spectrum used in CT requires a thick scintillator, because X-rays transmitted through the detector are wasted---they do not add to the measured signal, but contribute to the patient dose. Yet scintillation photons are emitted throughout this thickness and over a wide range of angles. Their lateral propagation broadens the detector point-spread function and preferentially suppresses the high spatial frequencies that encode fine anatomical structure. The efficiency–resolution trade-off shown in Fig.~1a reflects that the thickness required for dose-efficient detection is also what degrades the transfer of fine spatial detail.

\begin{figure*}
\centering
\vspace{-0.3cm}
  \includegraphics[scale=1]{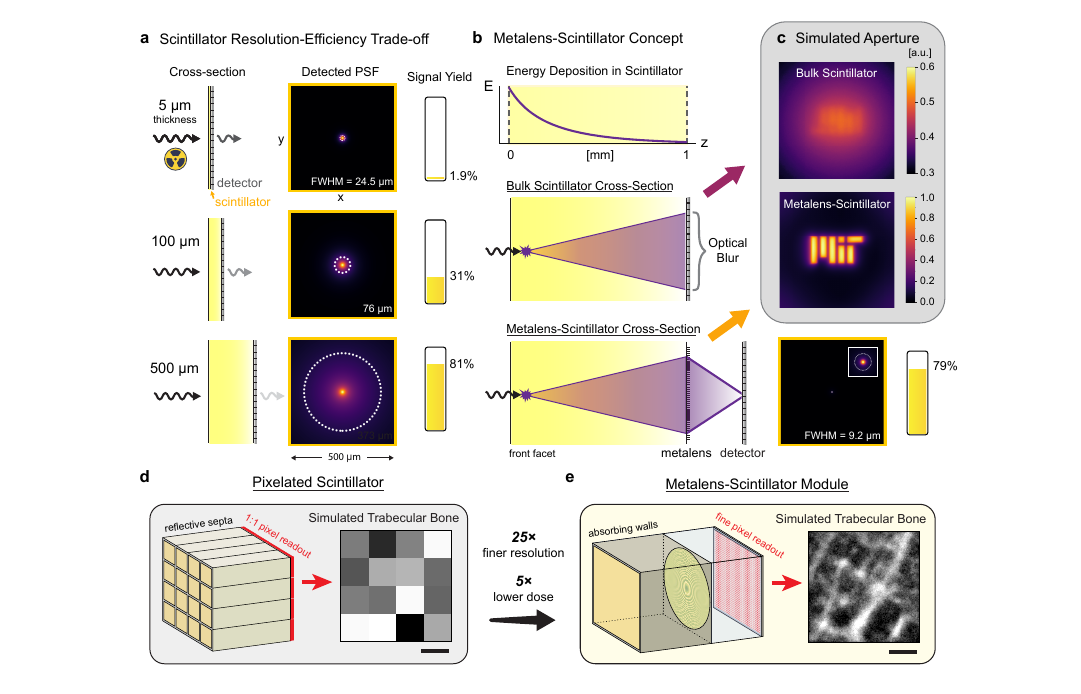}
  \caption{\small \textbf{The metalens-scintillator concept.} \textbf{a,} The scintillator thickness resolution-efficiency trade-off: increasing scintillator thickness improves X-ray absorption and signal yield, but degrades spatial resolution through lateral optical spreading. \textbf{b,} The metalens scintillator concept, simulated with a CdWO$_4$ scintillator irradiated by a standard 120 kVp X-ray spectrum, enabling both high resolution and signal yield. \textbf{c,} Simulated imaging of an ``MIT'' aperture for a bulk scintillator and the metalens-scintillator, using the scintillation angular spectrum framework outlined in Supplementary Note 1. Scale bars are 200 µm. \textbf{d,} Simulated imaging of trabecular bone using a pixelated scintillator detector common to energy-integrating CT. The fine trabecular structure shown lies below the spatial-resolution limit of conventional clinical CT and is therefore unresolved. \textbf{e,} Simulated imaging using a metalens-scintillator module for trabecular bone, showing enhanced resolution and dose efficiency for conditions relevant to full-body clinical CT. Scale bars in \textbf{d} and \textbf{e} are 0.5 mm.}
    \label{fig:1}
    \vspace{-0.3cm}
\end{figure*}

\begin{figure*}
\centering
  \includegraphics[width=183mm]{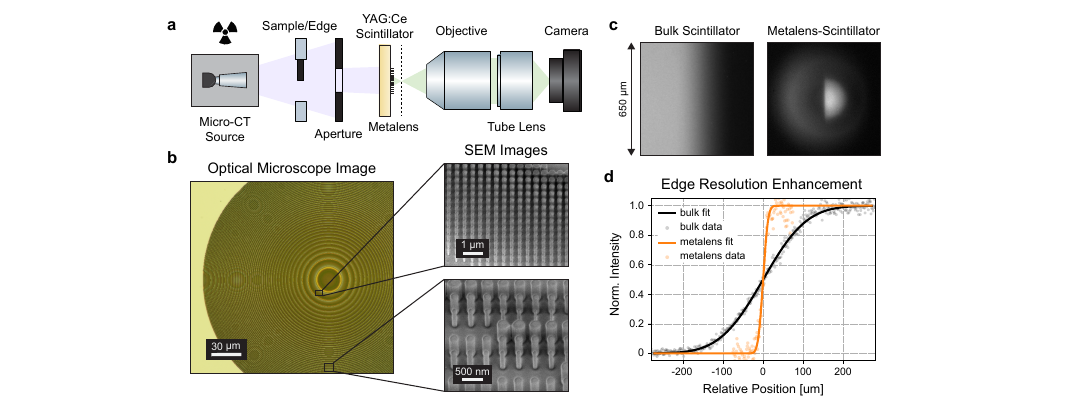}
  \caption{\small \textbf{Experimental setup, nanofabrication, and resolution measurements.} \textbf{a,} Experimental setup, where an objective lens focuses on either the back facet of the scintillator for the bulk scintillator, or a detector plane (dotted line) for the metalens-scintillator. \textbf{b,} Metalens-scintillator nanofabrication on YAG:Ce scintillator, showing an optical microscope image of the metalens, as well as SEM image insets showing silicon nitride nanopillar feature sizes down to 50 nm and 14$\times$ aspect ratio. \textbf{c,} Experimental images acquired from a sharp edge, showing large blur for the bulk scintillator, and a highly resolved edge for the metalens scintillator. The pattern is circular due to the aperture that restricts the X-ray illumination to the area of the metalens. \textbf{d,} Edge resolution fits and raw data plotted as points. The resolutions of the bulk scintillator and metalens-scintillator are 202.2 and 22.9 µm (Supplementary Note 6), respectively, showing 8.8$\times$ enhancement in resolution comparing the metalens-scintillator measurement to the bulk measurement in the experimental set-up.}
    \label{fig:2}
    \vspace{-0.3cm}
\end{figure*}

Existing detector architectures address this limitation by physically restricting optical spread via volumetric confinement and, in doing so, introduce trade-offs. Pixelated scintillator arrays use reflective septa to confine light, although their resolution is limited by pixel pitch and their active fill factor decreases as pixels become smaller~\cite{pelc2014recent,schwartz2025photon,baffour2022ultra}. Columnar CsI:Tl, phase-separated fibers, and filled micropores guide light through the absorbing volume, but can be limited by afterglow, optical loss, material choice, and the difficulty of volumetric fabrication~\cite{miller2005recent,ohashi2013submicron,chen2018simulated,yoshino2023development,hormozan2016high}. Direct-conversion semiconductors avoid optical transport and have enabled photon-counting CT, spectral imaging, and pixel sizes of approximately 150~µm~\cite{taguchi2013vision,willemink2018photon,mccollough2023technical,sartoretti2023photon,leng2018150,schwartz2025photon,leng2019photon}, but they introduce a different trade-off in which smaller pixels increase charge sharing and fluorescence crosstalk, while high X-ray fluxes cause pulse pile-up and count-rate nonlinearity~\cite{xu2011evaluation,hsieh2018spectral,persson2020detective,bhattarai2023exploration,tao2019feasibility,rajbhandary2020detective,mccollough2023technical}. Nanophotonic scintillators are an emerging approach to control the optical environment at the scale of the emission wavelength~\cite{roques2022framework,salomoni2018enhancing} where plasmonic~\cite{ye2024nanoplasmonic,bignell2013plasmonic,liu2017plasmonic}, photonic-crystal~\cite{roques2022framework,salomoni2018enhancing,kurman2020photonic,zhang2017enhanced,singh2018enhanced,Long2024NonreciprocalScintillation, martin2025large, Zabel2026Dosimetric}, inverse-designed~\cite{shultzman2023enhanced,min2025end,Chen2026WavefrontEngineering}, and volumetric~\cite{jurgensen2025volumetrically,Choi2026Supercollimating} structures have been proposed to enhance light yield, to accelerate emission through the Purcell effect~\cite{kurman2024purcell,roques2022framework}, or to control emission directionality~\cite{kurman2020photonic, shultzman2023enhanced, Choi2026Supercollimating}. Prior work, however, has largely optimized the number, timing, or emission direction of photons rather than considering the optical wavefront for spatial information transfer. This suggests a different strategy in scintillation-based imaging: rather than preserving resolution by volumetric confinement of scintillation light, the transfer of high-spatial-frequency information from within a thick scintillator to the detector can be nanophotonically controlled at the scintillator surface.

Here, we demonstrate this principle with a metalens fabricated directly on a bulk scintillator to transfer high-spatial-frequency information from within the scintillator volume to the detector. We implement this architecture using a silicon nitride metalens on a YAG:Ce scintillator and demonstrate imaging on inorganic and small biological specimens. Then, using clinical performance metrics and a model-based DQE and task-based detectability framework, we predict that the metalens-integrated detector can enable a 5$\times$ reduction in X-ray dose and a 25$\times$ increase in resolution bandwidth compared with pixelated scintillators used in state-of-the-art energy-integrating CT detectors. This approach helps to decouple X-ray absorption from optical image formation while retaining a planar geometry compatible with conventional scintillation-based systems.

\section{Metalens control of scintillation}

We first describe the metalens-scintillator concept, illustrated in \cref{fig:1}b. Under X-ray illumination, energy deposition in the scintillator follows Beer-Lambert attenuation, so a substantial fraction of the spatially modulated signal is generated near the entrance facet of the scintillator. In a bulk scintillator, the resulting optical emission spreads throughout the volume before detection, blurring high-spatial-frequency information. The metalens instead controls the propagation of this volumetrically generated scintillation light. The high flux of light generated near the scintillator front facet is transferred with high spatial fidelity, whereas weaker contributions generated near the back facet are increasingly defocused and contribute predominantly at lower spatial frequencies. High-spatial-frequency information can therefore be transferred to the detector without requiring all of the scintillation light to be sharply imaged. Wave-optical simulations of an aperture showing the logo of MIT in \cref{fig:1}c show that this metalens-scintillator readout produces a significantly higher-fidelity image than a bulk scintillator. A detailed framework for the scintillation wave-optics simulations is provided in Supplementary Note 1 and also in Ref.~\cite{Chen2026WavefrontEngineering}.

We also compare the metalens-scintillator to pixelated scintillators commonly used in clinical settings (\cref{fig:1}d,e) where the scintillator is divided into elements separated by reflective septa that confine scintillated light within each pixel for one-to-one photodetector readout. For the scintillator material and thickness, we reference the AcQSim CT detector due to the availability of its specifications, namely a 2.3-mm-thick CdWO$_4$ scintillator~\cite{garcia2002performance}, irradiated by a 120 kVp RQT 9 X-ray spectrum~\cite{ptb2015radiationqualities} commonly used in full-body CT. Furthermore, we model the pixelated scintillator with modern energy-integrating CT sampling: a 0.98-mm detector-plane pixel pitch~\cite{amsosram_as5951}, which corresponds to 0.52~mm when projected to scanner isocenter~\cite{baffour2022ultra,wang2018improving} using a standard geometric magnification~\cite{radiologykey_computed_tomography} of 1.9. The full simulation specifications, including gantry rotation speed, dosage, detector readout-noise, and more, are detailed in Supplementary Note 2. Additional simulations showing the signal-efficiency penalty incurred when pixel pitch is reduced at fixed pixelated septa width for a pixelated scintillator are provided in Supplementary Fig. 2. ~\cref{fig:1}e shows the metalens-scintillator module that can allow for large-area detection. We note that this design assumes absorptive walls and therefore a conservative 26.7\% light collection, as compared to 100\% light collection in the idealized pixelated scintillator model, also explained in Supplementary Note 2. This design choice was made to isolate the metalens-scintillator enhancement separately from additional considerations like cross-talk and reconstruction. Therefore, advanced methods including metalens arrays and image retrieval where significantly more signal is utilized could further improve detection fidelity. Even under this conservative assumption, the metalens-scintillator retains performance that is lost by the pixelated scintillator, producing a 25$\times$ increase in resolution bandwidth and a 5$\times$ lower required X-ray dose, as further quantified in the detector-level analyses below in \cref{fig:4}. This enhancement results from nanophotonic control of volumetrically generated light, which preserves and routes high-spatial-frequency information within a compact, high-NA architecture. Achieving the same NA at the same working distance with a conventional refractive lens would require a bulk material with a refractive index exceeding 5---well beyond that of available transparent optical materials. Realistic refractive designs are therefore geometrically restricted to $\mathrm{NA}\approx 0.45$, reducing their task-based performance below that of a standard pixelated scintillator (Supplementary Note~3).

\begin{figure*}
\centering
\vspace{-0.3cm}
  \includegraphics[scale=1]{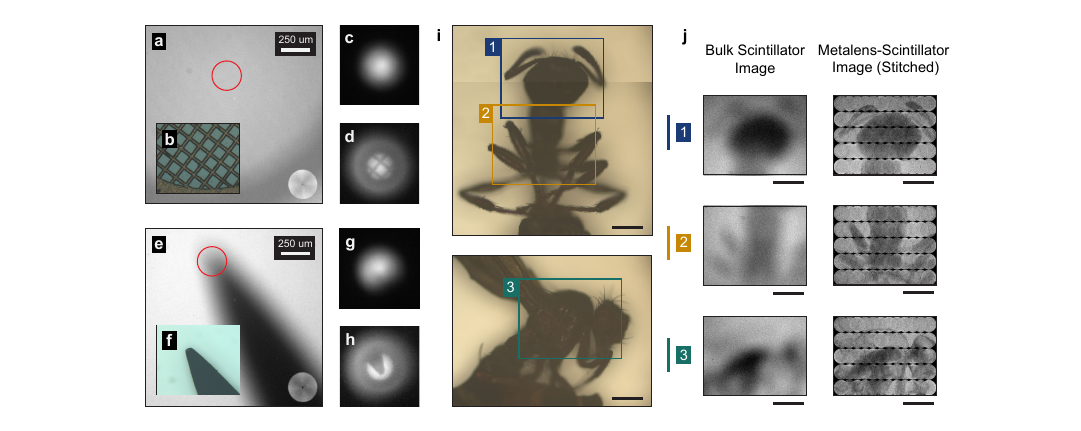}
  \caption{\small \textbf{Imaging inorganic and biological specimens} \textbf{a,} Bulk scintillator imaging of a TEM grid with a red circle indicating the region of interest imaged by the metalens-scintillator. The unused metalens can be seen in-focus on the bottom right of the image. \textbf{b,} Optical image of the TEM grid. \textbf{c,} Bulk scintillator imaging with an aperture for comparison to \textbf{d,} metalens-scintillator imaging showing enhanced resolution of TEM grid. \textbf{e-h,} Corresponding images of a needle pinpoint. The metalens-scintillator image is inverted by the metalens projection, but retains the high-spatial-frequency features of the object. \textbf{i,} Optical image of an iodine-stained ant (\textit{Tetramorium immigrans}) and fruit fly (\textit{Drosophila melanogaster}) with labeled regions. Iodine is used to enhance X-ray imaging contrast, with details in Supplemental Note 8. \textbf{j,} Comparison of bulk scintillator image and metalens-scintillator image for labeled regions. Since the region of interest is larger than the imaging area of the 250~\textmu m diameter metalens, the metalens-scintillator images are stitched from raster-scanned images, with processing and stitching details in Supplemental Note 7. Scale bars in \textbf{i} and \textbf{j} are 500~\textmu m.}
    \label{fig:3}
    \vspace{-0.3cm}
\end{figure*}

\section{Recovering fine spatial information}

We fabricated a silicon nitride metalens directly on a 1.0-mm-thick YAG:Ce scintillator that emits broadband wavelengths around a peak of 550 nm (\cref{fig:2}b). The metalens was designed using the local periodic approximation with NA = 0.45. Fabricating the metalens required transferring a nanoscale pattern of spatially varying nanopillar diameters through a 700-nm-thick silicon nitride layer while preserving the smallest features. We deposited low-stress silicon nitride by low-temperature plasma-enhanced chemical vapour deposition to promote adhesion to YAG:Ce, followed by 60~nm of chromium to provide a robust hard mask for the deep nitride etch. The nanopillar pattern was written in HSQ by electron-beam lithography, transferred into the chromium by Cl$_2$/O$_2$ ICP reactive-ion etching, and then transferred through the silicon nitride by CHF$_3$ ICP reactive-ion etching. Finally, the chromium mask was removed by wet etching, leaving the silicon nitride nanopillar metalens. Further nanofabrication and metalens design details are found in Supplementary Note 4. 

Under X-ray illumination from a ZEISS Xradia 620 Versa source, we compared two readout conditions: conventional bulk-scintillator imaging, in which the objective was focused at the back facet of the scintillator, and metalens-scintillator imaging, in which the objective was focused at the metalens image plane. We used 50 kVp in the micro-CT to mimic the energy deposition in the YAG:Ce scintillator to that of CdWO$_4$ when irradiated by the 120 kVp RQT 9 source, shown in Supplementary Note 5. Imaging a sharp edge shows an 8.8$\times$ enhancement in spatial resolution, shown in \cref{fig:2}c and d. Details for edge resolution measurements can be found in Supplementary Note 6. These experimental resolution measurements are predicted by the multi-wavelength, noise-aware scintillation imaging wave-optics simulation framework detailed in Supplementary Note 7, helping to validate the same scintillation-transport model later used for the detector-level dose and resolution analysis. 

\begin{figure*}
\centering
  \includegraphics[scale=1]{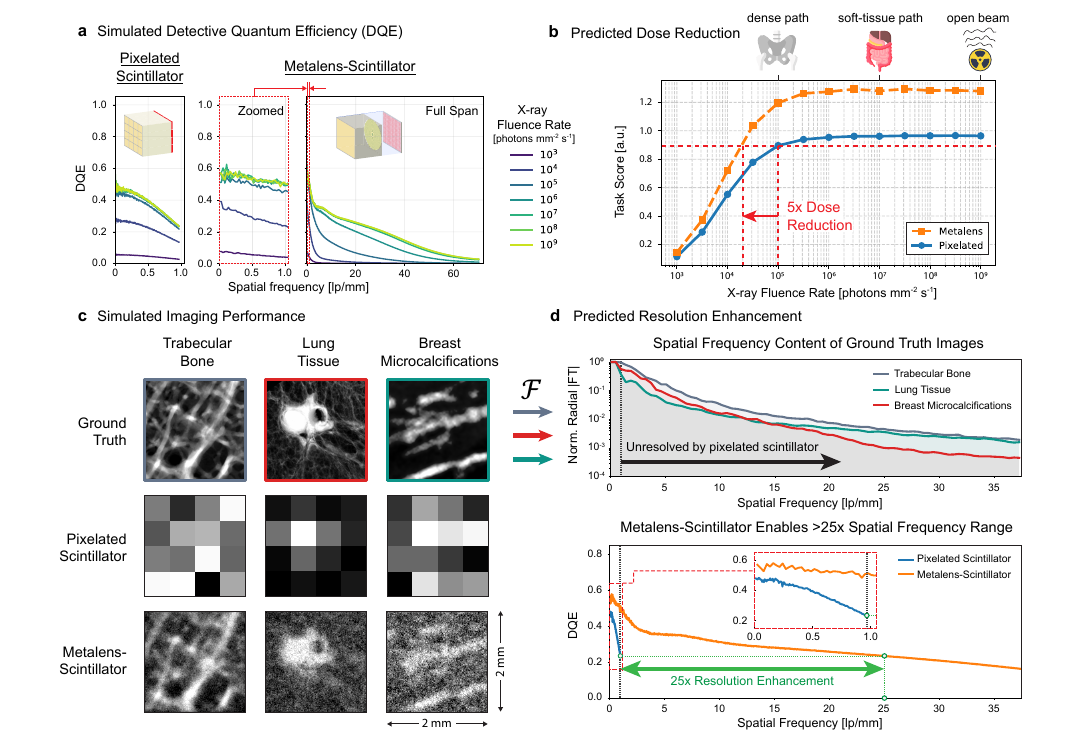}
  \caption{\small 
  \textbf{Predicted dose and resolution gains of the metalens-scintillator.} 
  \textbf{a,} Simulated detective quantum efficiency (DQE) for the pixelated scintillator and metalens-scintillator over X-ray fluence rates from $10^3$ to $10^9$ photons/mm$^{2}$/s are calculated using CT detector parameters described in Supplementary Note 2, and full details of the DQE calculation including the MTF, NNPS, and DQE calculation can be found in Supplementary Note 9. The metalens-scintillator preserves DQE to spatial frequencies approaching 70 lp mm$^{-1}$; the inset shows the low-frequency region for comparison with the pixelated-scintillator Nyquist frequency of 0.97 lp mm$^{-1}$, corresponding to a \(0.98\)-mm physical detector pitch, or \(0.52\) mm when projected to isocenter with a geometric magnification of 1.9.
  \textbf{b,} Task-based detectability score calculated from the DQE according to AAPM Report No. 233~\cite{samei2019performance}, with task profile and calculation details provided in Supplementary Note 10, demonstrating a projected 5$\times$ reduction in required X-ray fluence for the metalens-scintillator. 
  \textbf{c,} Simulated imaging of trabecular bone, lung tissue, and breast microcalcifications for ground-truth images, the pixelated scintillator imaging, and the metalens-scintillator imaging. Dataset and simulation details are provided in Supplementary Note 11. Monte-Carlo Geant4 particle-transport validation simulations are provided in Supplementary Note 12. 
  \textbf{d,} Frequency-domain comparison showing that substantial high-spatial-frequency content in the ground-truth images lies beyond the pixelated-scintillator Nyquist limit. Using the DQE at the pixelated-scintillator Nyquist frequency as a common performance threshold (dotted green lines), the metalens-scintillator reaches the same DQE value at a 25$\times$ higher spatial frequency, giving a projected 25$\times$ resolution-bandwidth enhancement; DQE calculations are again found in Supplementary Note 9.}
    \label{fig:4}
    \vspace{-0.3cm}
\end{figure*}

We next applied the metalens-scintillator to X-ray imaging of inorganic and biological specimens. For inorganic specimens, the bulk scintillator blurs the fine bars of a TEM grid and the sharp boundary of a needle pinhole, whereas the metalens-scintillator recovers the corresponding high-spatial-frequency features within the metalens field of view (\cref{fig:3}a-h). We then imaged iodine-stained biological specimens with fine internal and external structure, including an ant and fruit fly. Because the specimens exceeded the 250~\textmu m metalens imaging diameter, the metalens-scintillator image was formed by raster scanning and stitching adjacent circular fields of view. Compared with bulk scintillator images acquired under the same X-ray imaging conditions, the stitched metalens-scintillator images recover substantially finer anatomical detail (\cref{fig:3}i,j). For example, Region 3 shows clear iodine staining in the internal organs of the fruit fly. Additional information on sample preparation and metalens-scintillator stitching is found in Supplementary Note 8.

\section{Dose-efficient high resolution detection}

Having established close agreement between the simulated and experimental edge responses (Supplementary Note 7), we next used the same wave-optical transport framework to project detector-level performance in a CT-relevant geometry. We calculated the modulation transfer function (MTF), normalized noise power spectrum (NNPS), and detective quantum efficiency (DQE) for both the pixelated scintillator and metalens-scintillator across incident X-ray fluence rates spanning highly attenuated anatomical paths to open-beam conditions (\cref{fig:4}a). Comprehensive details of the calculation can be found in Supplemental Note 8. The pixelated scintillator provides high DQE only below its Nyquist frequency, whereas the metalens-scintillator maintains nonzero DQE over a much broader spatial-frequency range, reflecting its finer sampling and optically retrieved readout of the thick scintillator. Notably, this broader DQE bandwidth is maintained even with the metalens-scintillator light-collection efficiency limited to 26.7\% in the simulations.

To estimate the potential dose impact of this difference, we used the AAPM Report No. 233 task-based framework~\cite{samei2019performance} to calculate a detectability score from the DQE for a spatial task (3 mm "Designer Profile") whose frequency content notably lies fully below the pixelated-scintillator Nyquist frequency. Details of the AAPM task-based framework implementation are explained thoroughly in Supplementary Note 10. Across the simulated fluence range, the metalens-scintillator reaches the same task score at approximately fivefold lower incident X-ray fluence than the pixelated scintillator (\cref{fig:4}b). Thus, for tasks already resolvable by conventional energy-integrating detectors (below its Nyquist frequency), the projected benefit of metalens-scintillator readout is also improved dose efficiency. We use a fluence rate of 10$^5$ as a reference because this is the rate at which the detection ability of the pixelated detector begins to degrade. This value of 10$^5$ is not a coincidence: standard CT uses an open-beam fluence rate of 10$^9$ such that the transmitted fluence rate through the densest regions of interest (such as thick bone) is 10$^5$, and can still be properly detected~\cite{taguchi2013vision,danielsson2021photon}. Therefore, the metalens-scintillator equivalent task score at 2 $\times$ 10$^4$ fluence rate is an indication that the same detectability can be achieved at this lower rate of X-rays incident at the detector. Equivalently, the open-beam fluence can be lowered fivefold, lowering overall X-ray dosage to the patient. 

We then assessed tasks with finer spatial content using simulated images of trabecular bone, lung tissue, and breast microcalcifications (\cref{fig:4}c). Resolving such fine structures can provide clinically important diagnostic information. For example, trabecular architecture can improve fracture-risk assessment, fine pulmonary features can support lung-nodule characterization, and microcalcifications can indicate early breast cancer~\cite{samelson2019cortical,yan2025low,rauch2016microcalcifications,suryanarayanan2007detection}, among many other clinical applications. These structures contain substantial spatial-frequency content beyond the pixelated-scintillator Nyquist limit, leading to severe loss of fine detail in the pixelated detector image. The simulation implementation, including datasets and material attenuation constants, are found in Supplementary Note 11. Furthermore, we validate the same metalens-scintillator architecture using Geant4 Monte-Carlo particle-transport simulations~\cite{agostinelli2003geant4}, a standard toolkit for modeling radiation interactions in scintillators. Geant4 tracks X-rays through the detector geometry using a comprehensive electromagnetic interaction model, including photoelectric absorption and secondary-electron generation, Rayleigh scattering, Compton scattering, and subsequent energy deposition in the scintillator. Each optical photon is then simulated with ray-tracing through the metalens-scintillator to the detector. Supplementary Note 12 provides the Geant4 implementation details and shows close agreement with the fast statistical scintillation framework described in Supplementary Notes 9 and 11. The frequency-domain interpretation in \cref{fig:4}d shows that the metalens-scintillator preserves detector response over a substantially wider spatial-frequency range. Using the DQE at the pixelated-scintillator Nyquist frequency as a common performance threshold, the metalens-scintillator reaches the same DQE value at a 25$\times$ higher spatial frequency, giving a projected 25$\times$ resolution-bandwidth enhancement. Considering the full simulated metalens-scintillator bandwidth, which extends to 70 lp/mm, the accessible spatial-frequency span is 72$\times$ larger than that of the pixelated scintillator. Together, these simulations predict that metalens-based scintillator readout could improve both dose efficiency for conventional CT-scale tasks and spatial resolution for finer structures that are currently beyond the sampling limit of pixelated scintillator detectors.

\section{Outlook}

These results establish a different strategy for preserving spatial information in scintillation detectors. Instead of preserving resolution by segmenting, guiding, or structuring the full scintillator volume, high-spatial-frequency information can be transferred optically from a thick scintillator without requiring all of the generated scintillation light to remain spatially confined or sharply imaged. This separates, to a substantial degree, the material function of efficiently absorbing X-rays from the optical function of forming a high-resolution image. Experimentally, the device recovers fine spatial detail from a bulk scintillator. In CT-relevant detector simulations, the same experimentally validated transport model predicts improved dose efficiency for resolvable tasks and a substantially wider spatial-frequency bandwidth for finer structures.

Several avenues are envisioned to further enhance the performance of the metalens-scintillator device. The simulated detector module uses absorbing walls to suppress optical crosstalk, limiting collection efficiency to 26.7\%. Future designs could replace this simple isolation scheme with metalens arrays, larger-area fabrication, and reconstruction-based recovery of overlapping optical fields. Even under these conservative conditions, the projected gains show that nanophotonic control of scintillation can extend beyond light yield, emission rate, and directionality to the transfer of spatial information itself. More broadly, our results suggest a class of radiation detectors in which the scintillator and its optical interface together act not only as a radiation-to-light converter, but as an information-transfer system whose spatial response can be controlled.

\section{Author contributions}
J.C. and C.R.-C. initially conceived the project. 
J.C. performed nanofabrication with input from L.M.-M. and C.M.S.
J.C. built the experimental setup with input from S.P. and S.E.K.
J.C. and S.C. developed the theoretical and numerical models with input from W.M. 
J.C. analyzed the experimental data with input from S.P. and S.V. 
M.S., C.R.-C., J.H., and R.G. supervised the research. 
J.C., S.C., C.R.-C., and M.S. wrote the paper with inputs from all authors.

\section{Competing interests}
J.C., S.V., S.P., C.R.-C., and M.S. are pursuing patent protection for ideas in this work (U.S. Provisional Application No. 64/008,275).

\section{Data and code availability statement}
The code used to generate the simulations and analyses in this study is publicly available at \url{https://github.com/joshualchen/metalens-scintillator}. The data supporting the findings of this study are available from the corresponding authors upon reasonable request. Correspondence and requests for materials should be addressed to \mbox{chenjosh@mit.edu}.

\section{Acknowledgments}
The authors would like to thank William Li and Dr. Ryan Adams, MD for useful discussions, and Dina Volfson from the Littleton Lab for fruit flies. We would also like to thank staff at MIT.Nano: Mark Mondol, Cody Corey, Maansi Patel, Kurt Broderick, Eric Lim, Jim Daley, Gary Riggott, Donal Jamieson, and Bob Bicchieri. The authors acknowledge the MIT SuperCloud and Lincoln Laboratory Supercomputing Center for providing (HPC, database, consultation) resources that have contributed to the research results reported within this paper/report. OpenAI’s ChatGPT/Codex was used to assist with language editing, organization, and refinement of explanatory text during manuscript preparation. All scientific content and conclusions are those of the authors.

\section{Funding}
This material is based upon work supported in part by the U.S. Army Research Office through the Institute for Soldier Nanotechnologies at MIT under Cooperative Agreement Number W911NF-23-2-0121. J.C. and W.M. acknowledge support from the NSF GRFP under Grant Number 2141064. S.P. acknowledges support from the MathWorks Engineering Fellowship via MIT. S.~C.~acknowledges support from the Korea Foundation for Advanced Studies Overseas PhD Scholarship. L.M.M.
was supported by the DARPA ENvision program and a fellowship
from the Swiss National Science Foundation (P500PT-203222). C.R.-C. acknowledges startup funding from the Institute of Science and Technology, Austria (ISTA).  

\bibliographystyle{ieeetr}
\bibliography{bibliography}
\end{document}


\rmfamily




\title{Supplementary Information for: \\ Nanophotonic control of spatial information in scintillation detectors}

\author{Joshua~Chen}
\email{chenjosh@mit.edu}
\affiliation{Research Laboratory of Electronics, MIT, Cambridge, MA 02139, USA}

\author{Simo~Pajovic}
\affiliation{Department of Mechanical Engineering, MIT, Cambridge, MA 02139, USA}

\author{Seou~Choi}
\affiliation{Research Laboratory of Electronics, MIT, Cambridge, MA 02139, USA}

\author{Sachin~Vaidya}
\affiliation{Research Laboratory of Electronics, MIT, Cambridge, MA 02139, USA}
\affiliation{Department of Physics, MIT, Cambridge, MA 02139, USA}

\author{William~Michaels}
\affiliation{Research Laboratory of Electronics, MIT, Cambridge, MA 02139, USA}

\author{Louis~Martin-Monier}
\affiliation{Department of Materials Science and Engineering, MIT, Cambridge, MA 02139, USA}

\author{Christina~M.~Spägele}
\affiliation{Harvard John A. Paulson SEAS, Harvard University, Cambridge, MA 02138, USA}

\author{Steven~E.~Kooi}
\affiliation{Institute for Soldier Nanotechnologies, Massachusetts Institute of Technology, Cambridge, MA 02139, USA}

\author{Juejun~Hu}
\affiliation{Department of Materials Science and Engineering, MIT, Cambridge, MA 02139, USA}

\author{Rajiv~Gupta}
\affiliation{Neuroradiology Division, Department of Radiology, Massachusetts General Hospital, Harvard Medical School, Boston, MA 02114, USA}

\author{Charles~Roques-Carmes}
\affiliation{Institute of Science and Technology Austria (ISTA), Klosterneuburg 3400, Austria}

\author{Marin~Soljacic}
\affiliation{Research Laboratory of Electronics, MIT, Cambridge, MA 02139, USA}
\affiliation{Department of Physics, MIT, Cambridge, MA 02139, USA}

\noindent	

\noindent

\clearpage
\newpage

\renewcommand{\sp}{\sigma_+}
\newcommand{\sm}{\sigma_-}

\setlength{\parindent}{0em}
\setlength{\parskip}{.5em}
\vspace*{-2em}


\maketitle

\tableofcontents
\clearpage
\section{Scintillation angular spectrum simulation framework}

\begin{minipage}{\textwidth}
    \centering
    \includegraphics[width=0.9\textwidth]{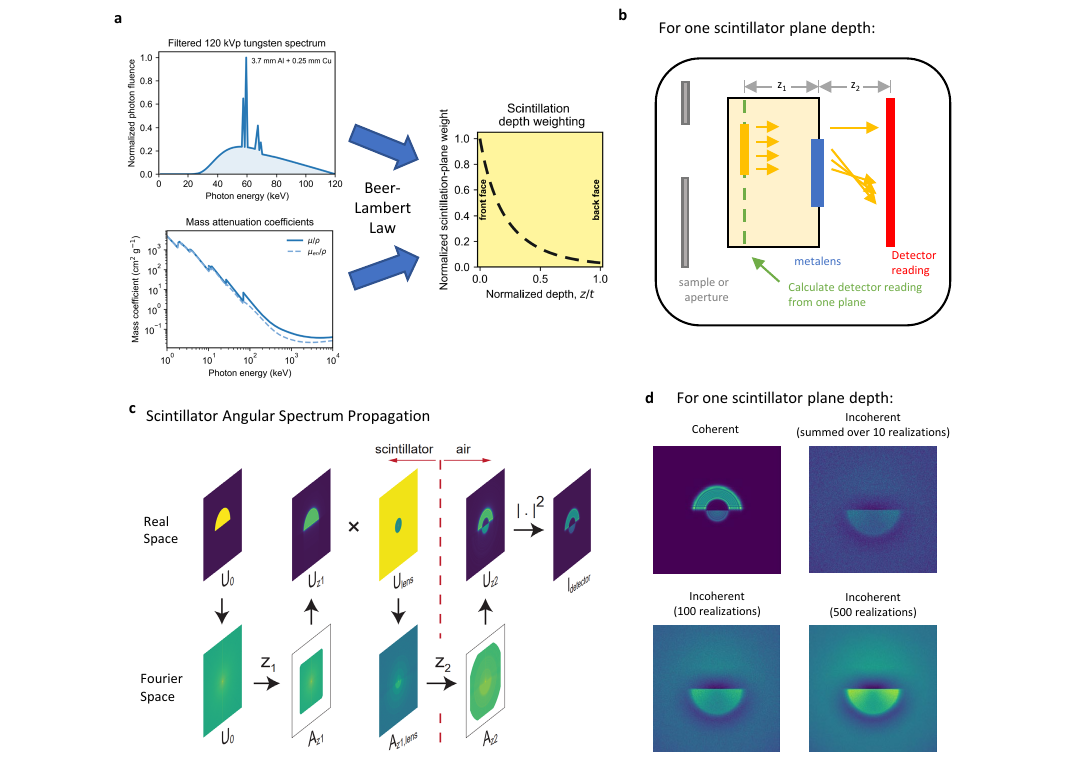}

    \captionof{figure}{\small
    \textbf{The scintillation angular spectrum simulation framework.}
    \textbf{a,} The X-ray source spectrum and the scintillator mass attenuation coefficients are combined to calculate the scintillator energy deposition according to the Beer-Lambert law.
    \textbf{b,} For a single discrete scintillator plane depth, a light field can be propagated through the scintillator by a distance of $z_1$ to the scintillator back facet where a metalens or arbitrary metasurface is fabricated, and then propagated a distance of $z_2$ to the detector plane. 
    \textbf{c,} The propagation of light from a single plane in the scintillator is calculated using the bandlimited angular spectrum method~\cite{matsushima2009band} where propagation occurs first through the scintillator, and then through air or another higher index medium to the detector. The fields are zero-padded to ensure there are no wrap-around artifacts, and sampled high enough in both the spatial and frequency domains to ensure no aliasing.
    \textbf{d,} Coherent propagation from a single scintillator plane results in a coherent detector reading. Since scintillation emission is incoherent, the incoherent image, again for a single scintillator plane, is calculated by employing a Monte-Carlo random phase averaging approach, where detector readings are averaged over many random phase realizations. More realizations results in a higher signal-to-noise ratio in the detector image. Ultimately, the total scintillation image is the energy-deposition-weighted (from part \textbf{a}) sum of all of the incoherent detector images from each scintillator depth.} 
    \label{fig:1}
\end{minipage}

\paragraph{Depth-dependent energy deposition.}
The scintillator is discretized into planes, with depth $x$ measured from the X-ray-incident face. For an incident spectral photon fluence $\Phi_0(E)$, the energy deposited in a slice extending from $x_i$ to $x_i+\Delta x_i$ is

\begin{equation}
W_i =
\int
\Phi_0(E)E
\frac{\mu_{\mathrm{en}}(E)}{\mu(E)}
e^{-\mu(E)x_i}
\left[1-e^{-\mu(E)\Delta x_i}\right]
\,\mathrm{d}E ,
\label{eq:depth_weight}
\end{equation}

where $\mu(E)$ and $\mu_{\mathrm{en}}(E)$ are the linear attenuation and energy-absorption coefficients, respectively. The exponential describes attenuation before the X-rays reach the slice, while the remaining terms describe the energy absorbed within it. The optical weight assigned to each plane is $w_i=W_i/\sum_j W_j$. This model assumes local energy deposition and does not consider secondary electrons or characteristic X-rays between slices. At diagnostic X-ray energies, secondary-electron ranges in dense scintillators vary from approximately $1~\mu$m at low energies to several tens of micrometers near 100~keV, remaining on the order of the detector sampling scale and therefore producing limited, localized spatial redistribution~\cite{martin2006recent}, see Supplemental Note 12.

\paragraph{Angular-spectrum propagation.}
Optical propagation from each scintillator plane is calculated using the band-limited angular spectrum method~\cite{matsushima2009band}. For a field $U(x,y)$ propagating a distance $z$ through a medium of refractive index $n$,

\begin{equation}
\mathcal{P}_{n,z}\{U\}
=
\mathcal{F}^{-1}
\left[
\mathcal{F}\{U\}
\exp\left(
i2\pi z
\sqrt{
\left(\frac{n}{\lambda_0}\right)^2-f_x^2-f_y^2
}
\right)
\right],
\label{eq:angular_spectrum}
\end{equation}

where $\lambda_0$ is the vacuum wavelength and $(f_x,f_y)$ are spatial frequencies. The angular spectrum is band-limited to prevent aliasing, and the fields are zero-padded to suppress FFT wrap-around artifacts.

For a source plane at depth $x_i$ in a scintillator of thickness $t$, the field first propagates a distance $z_{1,i}=t-x_i$ through the scintillator. At the back facet, it is multiplied by the complex transfer function $T(x,y)$ of the metalens, metasurface, or detector architecture and then propagates a distance $z_2$ through the detector-side medium:

\begin{equation}
U^{(\mathrm{det})}_{i,m}
=
\mathcal{P}_{n_2,z_2}
\left\{
T(x,y)
\mathcal{P}_{n_{\mathrm{scint}},z_{1,i}}
\left[
U^{(0)}_{i,m}(x,y)
\right]
\right\}.
\label{eq:two_stage_propagation}
\end{equation}

The same propagation framework is used for point sources, which generate depth-dependent PSFs, and for spatially extended apertures. These calculations are performed on the MIT SuperCloud high-performance computing system~\cite{reuther2018interactive}, using NVIDIA Volta V100 GPUs with 32 GB of memory on compute nodes equipped with Intel Xeon Gold 6248 processors.

\paragraph{Incoherent scintillation.}
Scintillation is modeled as spatially incoherent emission by assigning an independent random phase $\theta_m(x,y)\in[0,2\pi)$ to each realization:

\begin{equation}
U^{(0)}_{i,m}(x,y)
=
A_i(x,y)e^{i\theta_m(x,y)}.
\end{equation}

The incoherent response from depth $x_i$ is obtained by averaging detector intensities, rather than complex fields, over $N_{\mathrm r}$ independent phase realizations:

\begin{equation}
I_i(x,y)
=
\frac{1}{N_{\mathrm r}}
\sum_{m=1}^{N_{\mathrm r}}
\left|U^{(\mathrm{det})}_{i,m}(x,y)\right|^2.
\label{eq:phase_average}
\end{equation}

Increasing $N_{\mathrm r}$ reduces residual fluctuations from the finite random-phase ensemble, as shown in Fig.~S1d. Because emission from different depths is also mutually incoherent, the final detector image is the energy-deposition-weighted sum

\begin{equation}
I_{\mathrm{det}}(x,y)
=
\sum_i w_i I_i(x,y).
\label{eq:depth_sum}
\end{equation}

This framework describes both depth-dependent PSFs and arbitrary transmitting apertures for conventional and metalens-integrated scintillator architectures.

\clearpage
\section{Metalens--scintillator module full clinical simulation specifications}
\label{sec:clinical_simulation}

We compare a conventional pixelated scintillator detector with the proposed metalens--scintillator architecture under the same clinical conditions. The selected parameters are intended to represent a realistic full-body clinical CT system while allowing the architectures to be compared fairly.

\paragraph{Clinical CT reference.}
The simulations use the AcQSim 85-cm-bore CT scanner (Marconi Medical Systems, Inc., Cleveland, OH) as the clinical reference because its detector construction is reported in the literature~\cite{garcia2002performance}. The system employs CdWO$_4$ detector elements with a thickness of 2.3~mm along the X-ray propagation direction. We therefore use a 2.3-mm-thick CdWO$_4$ scintillator for both detector architectures. We note that CdWO$_4$ is a transparent scintillator that allows for the metalens-scintillator functionality; a translucent or opaque scintillator would severely degrade spatial information recovery.

\paragraph{X-ray spectrum and CdWO$_4$ properties.}
The incident spectrum follows the RQT~9 radiation quality defined for computed tomography in IEC~61267, shown in~\cref{fig:1}a. It consists of a 120~kVp tungsten spectrum with 3.7~mm Al and 0.25~mm Cu total filtration~\cite{ptb2015radiationqualities}. In simulation, this spectrum is generated with SpekPy~\cite{poludniowski2021spekpy}.

The CdWO$_4$ density is taken to be 7.90~g\,cm$^{-3}$. The energy-dependent mass attenuation and mass energy-absorption coefficients, $\mu/\rho$ and $\mu_{\mathrm{en}}/\rho$, respectively, are calculated for stoichiometric CdWO$_4$ using NIST attenuation-coefficient reference data~\cite{hubbell2004xray}. $\mu_{\mathrm{en}}/\rho$ determines the fraction of interacting X-ray energy deposited locally. We assume a CdWO$_4$ scintillation yield of 28 optical photons per keV of deposited X-ray energy~\cite{michail2020luminescence}. The energy-deposition calculation is identical for the pixelated and metalens--scintillator architectures. 

\paragraph{Incident fluence and temporal sampling.}
To represent high-speed clinical CT acquisition, we assume a gantry rotation time of 0.33~s and 1,000 projections per rotation~\cite{flohr2006developments}, corresponding to 0.33~ms per projection. Published estimates place the unattenuated fluence rate of a 120-kVp clinical CT beam at up to $10^9$ photons\,mm$^{-2}$\,s$^{-1}$, decreasing to approximately $10^5$--$10^8$ photons\,mm$^{-2}$\,s$^{-1}$ after bowtie-filter and patient attenuation~\cite{taguchi2013vision,danielsson2021photon}. We therefore consider dose rates from $10^3$ to $10^9$ photons\,mm$^{-2}$\,s$^{-1}$. The $10^9$ upper bound represents an unattenuated beam, while $10^5$ represents a strongly attenuated clinical projection path. Lower rates are included to characterize the transition into the photon-starved, read-noise-limited regime and to evaluate dose reductions below conventional operation.

\paragraph{Conventional pixelated scintillator electronics reference.}
In detective quantum efficiency (DQE) calculations, detector read noise must be considered for proper dose-dependent calculations. We therefore use representative commercial detector electronics to define the read-noise and sampling parameters for each architecture.

The conventional detector is modeled after the ams OSRAM AS5951 CT sensor~\cite{amsosram_as5951}, which has a 0.98~mm $\times$ 0.98~mm detector pitch and an input-referred noise of 0.30~fC, equivalent to $1.87\times10^3$ electrons RMS. Reflective separators in conventional clinical CT scintillator arrays are generally 100--200~$\mu$m wide, with approximately 100~$\mu$m representing the minimum practical width in state-of-the-art designs~\cite{hsieh2026nextgeneration}. We assume 100-$\mu$m-wide separators, giving an active width of 0.88~mm and a geometric fill factor of $\eta_{\mathrm{fill}}=[(0.98-0.10)/0.98]^2=0.806$.  The incident X-ray fluence is multiplied by this fill factor. More generally, reducing the detector pitch while retaining a fixed 100-$\mu$m separator width increases the inactive-area fraction and progressively reduces DQE, as shown in \cref{fig:pixel_pitch_dqe}. The separators are treated as perfectly reflecting, and optical cross-talk is neglected, such that 100\% of the light generated within the active scintillator area is collected. This provides an optimistic conventional-detector reference.

\begin{minipage}{\textwidth}
    \centering
    \includegraphics[width=0.8\textwidth]
    {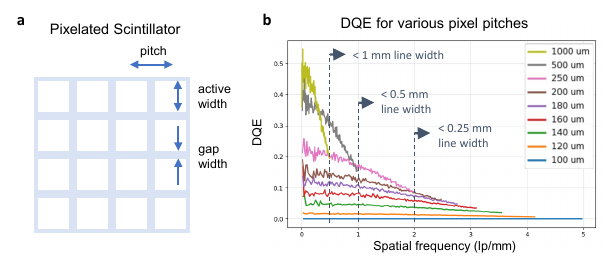}

    \captionof{figure}{\small
    \textbf{DQE penalty from reducing pixelated-scintillator pitch at fixed separator width.}
    \textbf{a,} Schematic of a pixelated scintillator array, showing the detector pitch, active scintillator width, and inter-element separator width.
    \textbf{b,} Simulated DQE for pixel pitches from 100 to 1,000~$\mu$m while holding the separator width fixed at 100~$\mu$m. Reducing the pitch increases the fraction of detector area occupied by inactive separators, and achieves a greater Nyquist frequency, but correspondingly reduces DQE. The indicated spatial frequencies correspond to line widths of 1, 0.5, and 0.25~mm for a line-and-space pattern.}
    \label{fig:pixel_pitch_dqe}
\end{minipage}

\paragraph{Metalens--scintillator architecture electronics reference.}
The metalens--scintillator detector is modeled using the commercially available Phantom v2640 high-speed CMOS camera as a realistic performance reference. At full resolution, this sensor combines a 13.5~$\mu$m pixel pitch, a read noise of 7.2 electrons RMS, and an acquisition rate of 4,855 frames per second~\cite{phantom_v2640_manual}. This exceeds the nominal acquisition rate of approximately 3,030 projections per second required for 1,000 projections during a 0.33-s CT rotation. We therefore assume 7.2 electrons RMS rather than the lower 1--2-electron noise achievable with slower scientific CMOS cameras.

\paragraph{Metalens aperture and light collection.}

Recent advances in metalens design using advanced methods have enabled experimentally demonstrated numerical apertures of up to 0.95--0.99~\cite{liang2019high,paniagua2018metalens,liang2018ultrahigh}. We therefore choose NA$=0.9$ as a conservative design value for the metalens-scintillator. The metalens is fabricated on the high-index CdWO$_4$ surface, and its NA is specified using its free-space focal-length. For the focal length associated with the 2.3-mm scintillator, this corresponds to a circular metalens diameter of 4.13~mm, which is the physical aperture used in the simulations. Light that does not intersect this aperture is treated as lost, and the module sidewalls are assumed to be perfectly absorbing to prevent optical cross-talk, but results in lower light collection efficiency, calculated as the following.

For isotropic emission at distance $d$ from a metalens of radius $r$, the directly collected fraction is calculated from the solid angle subtended by the aperture:
\begin{equation}
\eta_{\mathrm{dir}}(d)
=
\frac{1-\cos\!\left[\tan^{-1}(r/d)\right]}{2}.
\end{equation}
This depth-dependent collection fraction is used to scale the PSF from each scintillator plane before the energy-deposition-weighted depth summation. The resulting total direct-light collection efficiency here is 15.6\%.

Because scintillation emission can be modeled to be isotropic, a substantial fraction of the generated light initially propagates toward the X-ray-incident surface and away from the metalens. To recover part of this otherwise lost light, we also consider a mirror-assisted configuration in which an optically reflective, X-ray-transmitting coating is placed on the X-ray-incident scintillator surface, opposite the metalens. The mirror redirects backward-emitted scintillation light toward the metalens aperture. For an emitter located a distance $d$ from the metalens, the reflected path is equivalent to propagation from a virtual source at distance $2t-d$, giving
\begin{equation}
\eta_{\mathrm{mirror}}(d)
=
\eta_{\mathrm{dir}}(d)
+
\frac{1-\cos\!\left[\tan^{-1}\!\left(r/(2t-d)\right)\right]}{2},
\end{equation}
where $t=2.3$~mm is the scintillator thickness, resulting in a depth-weighted collection efficiency of 26.7\%. 

\paragraph{Common sensor and CT geometry assumptions.}
A quantum efficiency of 94\% is assumed for both detector architectures. The CT geometry assumed uses a representative source-to-isocenter distance (SID) of 50~cm and source-to-detector distance (SDD) of 95~cm~\cite{radiologykey_computed_tomography}, giving a geometric magnification of $M=\mathrm{SDD}/\mathrm{SID}=1.9$. Detector-plane dimensions are mapped to isocenter by division by $M$, while spatial frequencies are mapped according to $f_{\mathrm{iso}}=M f_{\mathrm{det}}$. Thus, the 0.98-mm detector pitch corresponds to approximately 0.52~mm at isocenter, representative of the approximately 0.5--0.625~mm detector sampling used in state-of-the-art energy-integrating CT systems~\cite{baffour2022ultra}.

\clearpage
\section{Refractive-lens constraints in a compact high-NA scintillator architecture}

\begin{minipage}{\textwidth}
    \centering
    \includegraphics[width=0.7\textwidth]{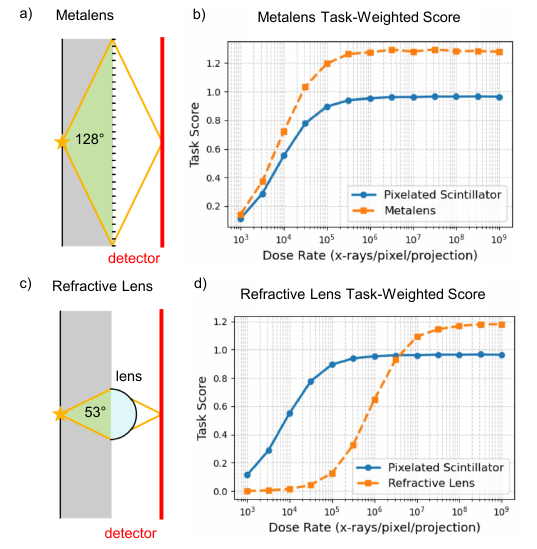}

    \captionof{figure}{\small
    \textbf{Geometric and performance comparison of scintillator-coupled imaging architectures.}
    \textbf{a,} Schematic diagram of the high-NA ($NA = 0.90$) monolithically integrated metalens-scintillator module.
    \textbf{b,} Task-weighted score versus dose rate for the metalens-scintillator system (see Supplemental Note 10 for information on the task-weighted detectability score).
    \textbf{c,} Schematic diagram of the geometrically restricted, low-NA ($NA = 0.45$) refractive lens configuration.
    \textbf{d,} Task-weighted score versus dose rate for the low-NA refractive lens configuration, demonstrating performance degradation below the standard pixelated scintillator baseline.}
    \label{fig:refractive_lens_comparison}
\end{minipage}

To understand why a nanophotonic metalens is required for the metalens-scintillator architecture, we must first evaluate the physical limits of replacing the metalens with a standard refractive lens. Our system assumes a 1x magnification metalens with a free-space Numerical Aperture (NA) of 0.9 (Fig.~\ref{fig:refractive_lens_comparison}a).

In a 1$\times$ magnification imaging configuration, the object distance ($z_1$) and image distance ($z_2$) are equal. By the thin lens equation, with $z_1 = z_2 = 1.0$ mm, the required focal length ($f$) is:
\begin{equation}
\frac{1}{z_1} + \frac{1}{z_2} = \frac{1}{f}
\implies
f = 0.5~\text{mm}.
\end{equation}

To achieve an NA of 0.9 at a working distance of $z_1 = 1.0$ mm, the lens must capture a half-angle ($\theta$) of:
\begin{equation}
\theta = \arcsin(\text{NA}) = \arcsin(0.9) \approx 64.16^\circ.
\end{equation}

This dictates the required physical diameter ($D_{\mathrm{req}}$) of the lens aperture:
\begin{equation}
D_{\mathrm{req}}
=
2z_1\tan(\theta)
=
2(1.0)\tan(64.16^\circ)
\approx
4.13~\text{mm}.
\end{equation}

Because standard refractive optics rely on volumetric phase accumulation, achieving this focal length requires aggressive physical curvature. Using the Lensmaker’s equation for a plano-convex lens, the required radius of curvature ($R$) is determined by the focal length and the material's refractive index ($n$):
\begin{equation}
\frac{1}{f} = \frac{n-1}{R}
\implies
R = f(n-1).
\end{equation}

As we push the curvature to bend light more steeply, we approach a hard physical limit: a lens cannot have a diameter larger than the sphere that defines its curvature. The absolute maximum physical diameter ($D_{\mathrm{max}}$) of a plano-convex lens is a perfect hemisphere, meaning the diameter is exactly twice the radius of curvature:
\begin{equation}
D_{\mathrm{max}} = 2R = 2f(n-1).
\end{equation}

For a lens to physically exist, its maximum allowed diameter must be greater than or equal to the required aperture ($D_{\mathrm{max}} \ge D_{\mathrm{req}}$). Substituting our system parameters ($f = 0.5$ mm, $D_{\mathrm{req}} = 4.13$ mm):
\begin{equation}
2(0.5)(n-1) \ge 4.13,
\end{equation}
\begin{equation}
n-1 \ge 4.13
\implies
n \ge 5.13.
\end{equation}

Therefore, to construct a single refractive lens capable of achieving an NA of 0.9 in this compact 1x relay architecture, the lens material must possess a refractive index greater than 5.13. No visibly transparent optical materials exist with an index this high.

\paragraph{Numerical aperture penalty and DQE degradation.}
If we construct the lens using an extremely high-index optical glass that is physically realizable, such as $n = 2.0$, we must significantly reduce the aperture for the lens to exist. Recalculating the maximum diameter for $n = 2.0$:
\begin{equation}
D_{\mathrm{max}}
=
2(0.5)(2.0-1)
=
1.0~\text{mm}.
\end{equation}

Restricting the lens diameter to 1.0 mm to satisfy the geometric constraints forces a significant reduction in the system's acceptance angle. The new restricted half-angle becomes $\theta = \arctan(0.5/1.0) = 26.56^\circ$, which yields a maximum achievable NA of 0.45 (Fig.~\ref{fig:refractive_lens_comparison}c).

This constrained acceptance angle severely degrades the light collection efficiency. When this NA penalty is propagated through the detective quantum efficiency (DQE) and task-weighted score calculation, the performance of the refractive system drops significantly below that of the standard pixelated scintillator baseline (Fig.~\ref{fig:refractive_lens_comparison}d). In contrast, the metalens-scintillator preserves its high-NA collection efficiency, yielding a superior task-weighted score (Fig.~\ref{fig:refractive_lens_comparison}b), precluding the use of a standard refractive architecture.

\paragraph{The Nanophotonic Advantage.}
Nanophotonics fundamentally circumvents the macroscopic limitations of bulk refraction. While metalenses are often recognized for reducing optical weight and size, their microscopically thin profile is a direct requirement for this architecture. Because a metalens relies on subwavelength diffraction rather than bulk refraction, it imparts the necessary phase delay abruptly at the emission interface. This decouples the optical power from the geometric thickness and curvature limits of bulk materials.

In addition to bypassing the refractive index limit, the planar metalens architecture provides several other advantages:

\begin{itemize}
    \item \textbf{Scalable monolithic integration:} X-ray and CT imaging require large, densely packed active areas. Assembling and aligning thousands of macroscopic lens elements across an array is highly complex for manufacturing. In contrast, metalenses can be monolithically patterned directly onto the scintillator wafer using standard semiconductor lithography, allowing for complete optical fill factor over large areas.

    \item \textbf{Optical aberration control:} The geometric limits derived above assume idealized ray tracing. In practice, pushing a single bulk refractive lens to a high NA introduces severe spherical and off-axis aberrations. These aberrations blur the point spread function (PSF) and further degrade the system's spatial resolution and task-based performance. In contrast, the high dimensional degrees of freedom of a metalens provide localized design control to significantly mitigate the severe geometric aberrations directly at the interface. This capability is enabled by extensive research in computational inverse design and full-wave simulations~\cite{molesky2018inverse,li2022inverse,pestourie2018inverse,li2022empowering}.

    \item \textbf{Mechanical stability on a CT gantry:} Full-body clinical CT gantries rotate at high speeds. A traditional lens array adds mass and would require a rigid housing to maintain optical alignment. The planar metalens adds negligible mass and is directly integrated with the pixel emission volume. Furthermore, direct patterning onto the scintillator removes the need for mechanical standoffs, alignment barrels, or indexing adhesives. This results in a consolidated device package that remains stable under the stresses of CT measurement.
\end{itemize}

Therefore, when traditional lens-coupled scintillation imaging systems are utilized, they are typically restricted to low-NA, dose-unrestricted applications such as micro-CT. The nanophotonic integration presented here enables the scaling of high-NA, high-efficiency optical routing for larger clinical systems.

\clearpage
\section{Design and nanofabrication of nitride-on-YAG:Ce metalens}

\paragraph{Nanofabrication of nitride-on-YAG:Ce metalens.}

YAG:Ce substrates (1~mm thick and 10~mm in diameter; Crytur) were cleaned sequentially with acetone and isopropyl alcohol. A 700-nm-thick low-stress silicon nitride film was deposited by low-temperature plasma-enhanced chemical vapor deposition, followed by 60~nm of chromium deposited by electron-beam evaporation. The low-stress nitride deposition was selected to promote adhesion to the YAG:Ce substrate.

Hydrogen silsesquioxane (HSQ) was used as the electron-beam resist for patterning the chromium hard mask. SurPass 4000 adhesion promoter was spin-coated at 3,000~r.p.m. for 60~s, followed by a 30-s isopropyl-alcohol rinse and a 30-min bake at 180\,$^{\circ}$C. A 6\% HSQ solution was subsequently spin-coated at 3,000~r.p.m. for 60~s. The metalens pattern was exposed by electron-beam lithography at an acceleration voltage of 125~kV and a beam current of 2~nA.

The exposed HSQ was developed through four sequential cycles, each comprising 15~s in a developer containing 1\% NaOH, 4\% NaCl, and 95\% water by mass, followed by a 15-s water rinse with gentle agitation. The sample was dried with N$_2$ after the final cycle. The developed HSQ pattern was transferred into the chromium layer by Cl$_2$/O$_2$ inductively coupled plasma reactive-ion etching (ICP-RIE). The chromium etch rate was measured by optical absorption spectroscopy. Residual CrCl$_x$ and CrO$_x$ were removed by a 1-min Ar sputter at 450~W, 5~mTorr, and 60\,$^{\circ}$C.

The chromium pattern was subsequently transferred through the 700-nm silicon nitride layer using CHF$_3$ ICP-RIE, with the chromium serving as the hard mask. After nitride etching, the remaining chromium was removed by a 75-s wet etch in Transene chromium etchant.

\paragraph{Metalens focal-length and image-plane selection.}
To select the nominal focal length associated with the desired object and image distances, we first consider the paraxial phase curvature
\begin{equation}
\phi_{\mathrm{in}}(r)
=
\frac{k_0 n_{\mathrm{YAG}}r^2}{2z_1},
\end{equation}
where $r$ is the radial coordinate, $k_0=2\pi/\lambda_0$, and $n_{\mathrm{YAG}}=1.82$. A metalens with nominal free-space focal length $f$ adds the phase
\begin{equation}
\phi_{\mathrm{ML}}(r)
=
-\frac{k_0r^2}{2f}.
\end{equation}
In the same paraxial approximation, a field converging to an image at a distance \(z_2\) in air has the phase curvature
\begin{equation}
\phi_{\mathrm{out}}(r)
=
-\frac{k_0r^2}{2z_2}.
\end{equation}
Equating these paraxial phase curvatures gives the refractive-index-weighted thin-lens equation
\begin{equation}
\frac{n_{\mathrm{YAG}}}{z_1}
+
\frac{1}{z_2}
=
\frac{1}{f}.
\label{eq:metalens_imaging}
\end{equation}

We select \(z_2=0.50\) mm as a convenient air-side image distance and \(z_1=0.91\) mm, such that \(z_1/n_{\mathrm{YAG}}=z_2\). This symmetric reduced-conjugate geometry gives approximately unit magnification and places the design plane \(90~\mu\mathrm{m}\) inside the X-ray-incident surface, within the region of substantial energy deposition while avoiding focus exactly at the surface. These parameters were selected as a practical imaging geometry rather than through numerical optimization of resolution. Substitution into Eq.~\ref{eq:metalens_imaging} gives \(f=0.25\) mm. The collection objective is focused on the corresponding image plane, located 0.50~mm beyond the patterned surface.

The fabricated metalens implements the corresponding spherical-wave phase profile at the design wavelength $\lambda_0=550$~nm,
\begin{equation}
\phi(r)
=
-k_0\left(\sqrt{r^2+f^2}-f\right)
\pmod{2\pi}.
\end{equation}
Because scintillation occurs throughout the YAG:Ce thickness, only the selected design plane is nominally in focus. The selected plane lies near the X-ray-incident surface, where a substantial fraction of the Beer--Lambert-weighted scintillation emission originates. Light from this region is concentrated onto relatively few detector pixels, producing a much higher signal per pixel and therefore dominating the localized image contrast. Emission from more distant depths is distributed over broader point-spread functions and contributes primarily as a diffuse background rather than a concentrated image signal.

\clearpage
\section{Scintillator energy deposition between 50 kVp for YAG:Ce and 120 kVp for CdWO$_4$}

\begin{minipage}{\textwidth}
    \centering
    \includegraphics[width=0.75\textwidth]{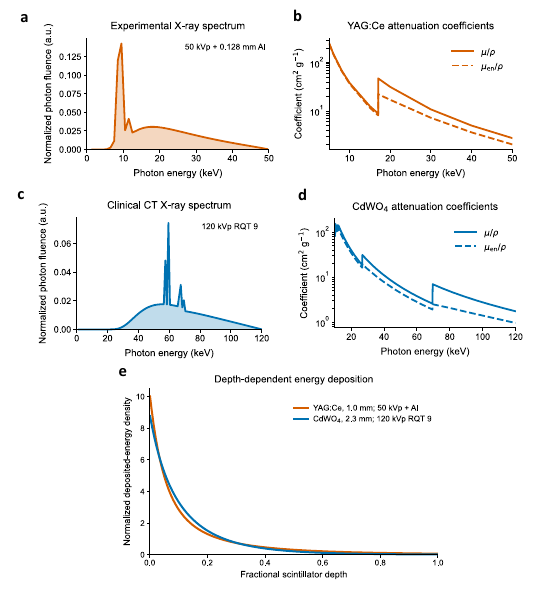}

    \captionof{figure}{\small
    \textbf{Experimental and clinical scintillator energy-deposition profiles.} The experimental source voltage and filtration are chosen to make the normalized depth-dependent energy-deposition profile as similar as practicable to that of the CdWO$_4$ clinical configuration.
    \textbf{a,} Simulated 50~kVp tungsten X-ray spectrum used in the experiment, filtered by eight 0.016-mm-thick aluminum-foil layers (0.128~mm total Al).
    \textbf{b,} Mass attenuation and mass energy-absorption coefficients calculated for YAG:Ce.
    \textbf{c,} The 120~kVp RQT~9 spectrum used in the clinical CT simulations, including 3.7~mm Al and 0.25~mm Cu filtration.
    \textbf{d,} Mass attenuation and mass energy-absorption coefficients calculated for CdWO$_4$.
    \textbf{e,} Normalized depth-dependent energy deposition in 1.0-mm-thick YAG:Ce under the experimental spectrum and 2.3-mm-thick CdWO$_4$ under the clinical spectrum. Depth is normalized by the respective scintillator thickness, with zero corresponding to the X-ray-incident surface.
    }
    \label{fig:yag_cdwo4_depth_match}
\end{minipage}

\paragraph{Experimental depth-profile matching.}
The proof-of-concept experiment uses a commercially available, 1.0-mm-thick YAG:Ce scintillator, selected for its availability in the required sample geometry and its well-established scintillation properties, whereas the clinical simulations use the 2.3-mm-thick CdWO$_4$ scintillator described in Supplementary Note~2. We select an experimental X-ray spectrum to reproduce the normalized depth distribution of energy deposition in the clinical configuration, enabling closer comparison despite different scintillator materials, thicknesses, and irradiation conditions. The experimental spectrum is generated for a 50~kVp tungsten source with eight layers of standard aluminum foil, each having a nominal thickness of 0.016~mm~\cite{vwr_reynolds_aluminum_foil}, giving 0.128~mm total Al filtration. The clinical spectrum is the 120~kVp RQT~9 spectrum used throughout the simulations and explained in Supplementary Note~2.

For both materials, the spectrum is combined with the corresponding $\mu/\rho$ and $\mu_{\mathrm{en}}/\rho$ coefficients using the Beer--Lambert energy-deposition model described in Supplementary Note~1. The coefficients are calculated from NIST elemental data using the stoichiometric compositions of Y$_3$Al$_5$O$_{12}$ and CdWO$_4$~\cite{hubbell2004xray}. The resulting profiles are expressed as a function of fractional scintillator depth, $u=x/t$, and normalized such that $\int_0^1 p(u)\,\mathrm{d}u=1$.

Thus, the experimental configuration is chosen to reproduce the relative depth distribution of scintillation emission in the clinical CdWO$_4$ configuration, enabling closer comparison between the two scintillator systems despite their different materials, thicknesses, and irradiation conditions.

\clearpage
\section{Scintillation straight-edge resolution measurement method}

\begin{minipage}{\textwidth}
    \centering
    \includegraphics[width=1\textwidth]{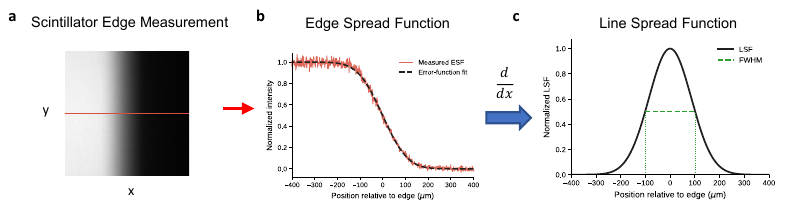}

    \captionof{figure}{\small
    \textbf{Scintillation straight-edge resolution measurement.}
    \textbf{a,} Experimental bulk-scintillator image of a straight X-ray-opaque edge. The intensity profile is extracted along the indicated line perpendicular to the edge.
    \textbf{b,} The resulting edge-spread function (ESF) and fitted error function.
    \textbf{c,} The line-spread function (LSF), calculated as the derivative of the fitted ESF. The LSF full width at half maximum is used as the spatial-resolution metric and is 202.2~$\mu$m for the measurement shown.}
    \label{fig:edge_resolution_method}
\end{minipage}

Simulation and experimental spatial resolution is determined from the image of a straight X-ray-opaque edge. An intensity profile is extracted perpendicular to the edge. This profile constitutes the edge-spread function (ESF). The measured ESF is fitted with an error function,
\begin{equation}
E(x)
=
a\,\mathrm{erf}\left(\frac{x-x_0}{c}\right)+d,
\end{equation}
where $a$ and $d$ describe the intensity range and background signal, respectively, $x_0$ is the edge position, and $c$ describes the transition width from high to low signal and vice versa. The line-spread function (LSF) is obtained by analytically differentiating the fitted ESF:
\begin{equation}
L(x)
=
\frac{\mathrm{d}E}{\mathrm{d}x}
=
\frac{2a}{\sqrt{\pi}c}
\exp\left[-\left(\frac{x-x_0}{c}\right)^2\right].
\end{equation}
The spatial resolution is therefore reported as the full width at half maximum of the LSF, $\mathrm{FWHM}=2\sqrt{\ln 2}\lvert c\rvert$, where $c$ is from fitting the analytical ESF to the experimental data.

\clearpage
\section{Comparison of simulation prediction and experimental measurements}

\begin{minipage}{\textwidth}
    \centering

    \makebox[\linewidth][c]{%
        \includegraphics[width=0.8\linewidth]{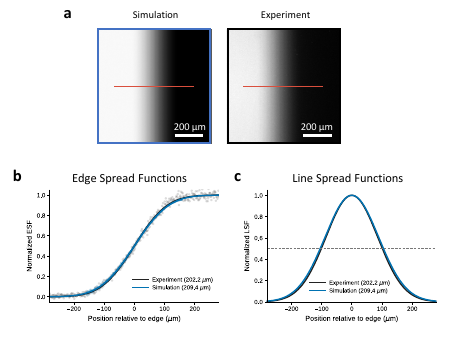}%
    }

    \captionof{figure}{\small
    \textbf{Comparison of simulated and experimental bulk-scintillator edge responses.}
    \textbf{a,} Simulated and experimental bulk-scintillator images. The red lines indicate the regions used to extract the edge profiles.
    \textbf{b,} Normalized bulk-scintillator edge-spread functions (ESFs) from simulation and experiment.
    \textbf{c,} Corresponding normalized line-spread functions (LSFs), giving FWHM values of 209.4~$\mu$m and 202.2~$\mu$m for the simulation and experiment, respectively.}
    \label{fig:bulk_simulation_experiment_comparison}
    \vspace{0.3cm}
\end{minipage}

We evaluate the scintillation wave-optics framework described in Supplementary Note~1 against the proof-of-concept experiment before analyzing it in the context of a CT detector geometry. The following comparison tests whether the modeled depth-dependent scintillation generation, multi-wavelength optical propagation, incoherent averaging, metasurface response, and experimental collection optics reproduce the measured bulk and metalens-scintillator edge responses.

The experimental comparison uses the 1.0-mm-thick YAG:Ce scintillator, the measured 50~kVp spectrum with eight layers of Al foil, or 0.128~mm Al filtration, and the corresponding depth-dependent energy-deposition weights described in Supplementary Note~1, emulating our experiments. Scintillation emission is calculated at 525, 550, 575, 600, and 625~nm and weighted by the YAG:Ce emission spectrum. At each wavelength and emission depth, spatial incoherence is modeled by averaging the detector intensities obtained from independent random-phase realizations. The experimental metalens geometry and propagation distances are retained.

\begin{minipage}{\textwidth}
    \centering

    \makebox[\linewidth][c]{%
        \includegraphics[width=0.8\linewidth]{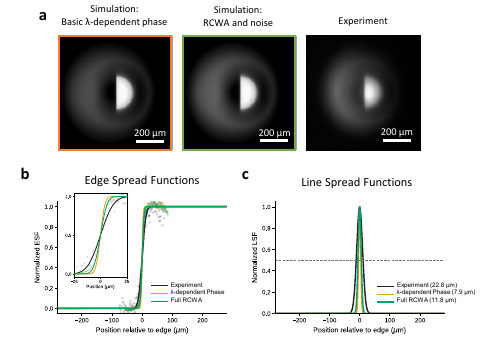}%
    }

    \captionof{figure}{\small
    \textbf{Comparison of simulated and experimental metalens--scintillator edge responses.}
    \textbf{a,} Simulated metalens--scintillator images obtained using the designed wavelength-dependent phase with uniform transmission, and the wavelength-dependent complex RCWA response, compared with the experimental image.
    \textbf{b,} Corresponding normalized edge-spread functions (ESFs). The inset shows the transition region over a narrower spatial range. The ESFs use the same normalization and axis ranges as the bulk-scintillator comparison in \cref{fig:bulk_simulation_experiment_comparison}b.
    \textbf{c,} Corresponding normalized line-spread functions (LSFs), giving FWHM values of 7.9~$\mu$m for the uniform-transmission model, 11.8~$\mu$m for the full-RCWA model, and 22.8~$\mu$m for the experiment. The LSFs use the same axis ranges as those in \cref{fig:bulk_simulation_experiment_comparison}c.}
    \label{fig:metalens_simulation_experiment_comparison}
    \vspace{0.3cm}
\end{minipage}

The collection optics are modeled using the specified $\mathrm{NA}=0.30$ of the Nikon 10$\times$ CFI60 TU Plan EPI objective used in the experiment~\cite{edmundoptics_nikon_cfi60}. For each random-phase realization, the objective collection NA is represented by a wavelength-dependent spatial-frequency-domain circular pupil applied to the frequency domain of the complex field,
\begin{equation}
P_{\lambda}(f_x,f_y)=
\begin{cases}
1, & \sqrt{f_x^2+f_y^2}\leq \mathrm{NA}/\lambda,\\
0, & \text{otherwise}.
\end{cases}
\end{equation}
The pupil is applied before calculating the detector intensity, after which the intensities are averaged over the incoherent realizations. This accounts for the finite spatial-frequency acceptance of the experimental objective within the wave-optical propagation model.

The bulk scintillator comparison isolates the components of the simulation that do not depend on the metasurface, including depth-dependent scintillation generation, propagation through YAG:Ce, incoherent phase averaging, spectral weighting, and finite-NA objective collection. The simulated and experimental images produce closely overlapping ESFs and LSFs (\cref{fig:bulk_simulation_experiment_comparison}a--c, with LSF FWHM values of 209.4~$\mu$m and 202.2~$\mu$m, respectively. This agreement supports the accuracy of the underlying scintillation-transport and collection model.

\begin{minipage}{\textwidth}
    \centering

    \makebox[\linewidth][c]{%
        \includegraphics[width=1\linewidth]{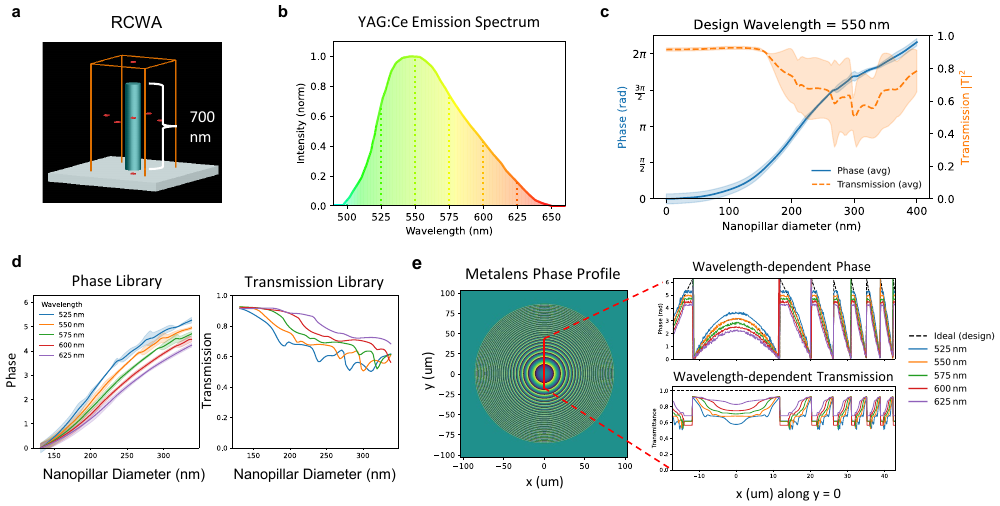}%
    }

    \captionof{figure}{\small
    \textbf{Wavelength-dependent RCWA model of the fabricated metalens.}
    \textbf{a,} Periodic RCWA unit cell consisting of a cylindrical silicon-nitride nanopillar with variable diameter, a fixed height of 700~nm, and a lattice period of 400~nm.
    \textbf{b,} YAG:Ce emission spectrum. Dotted lines indicate the five wavelengths used in the simulations.
    \textbf{c,} Nanopillar phase and transmission libraries at the 550-nm metalens design wavelength. Solid and dashed curves show the responses averaged over the simulated incidence-angle and polarization conditions, and the shaded regions show their corresponding variation.
    \textbf{d,} Phase and transmission libraries calculated at 525, 550, 575, 600, and 625~nm.
    \textbf{e,} Nanopillar-based metalens phase profile designed at 550~nm and cross-sections of the resulting wavelength-dependent phase and transmission. The same physical nanopillar-diameter map is evaluated at every wavelength; the dashed line indicates the ideal target phase or unity transmission.}
    \label{fig:rcwa_framework}
    \vspace{0.3cm}
\end{minipage}

The fabricated metalens is modeled using an RCWA framework summarized in \cref{fig:rcwa_framework}. The metasurface consists of cylindrical silicon-nitride nanopillars with a fixed height of 700~nm on a 400-nm-period lattice, with the nanopillar diameter controlling the transmitted phase and amplitude (\cref{fig:rcwa_framework}a). Because YAG:Ce emits over a broad spectrum (\cref{fig:rcwa_framework}b), the zero-order complex transmission is calculated as a function of nanopillar diameter at 525, 550, 575, 600, and 625~nm. At the 550-nm design wavelength, the phase and transmission are averaged over the simulated incidence-angle and polarization conditions, while the shaded regions indicate their variation within the range of incidence angles and both polarizations (\cref{fig:rcwa_framework}c). The target metalens phase is converted into a single physical nanopillar-diameter map using this 550-nm phase library, with the available phase range and nanopillar diameters restricted to those used in fabrication. The same diameter map is then evaluated using the phase and transmission libraries at each emission wavelength (\cref{fig:rcwa_framework}d), producing wavelength-dependent complex metalens responses that are weighted by the YAG:Ce emission spectrum and propagated through the scintillation imaging model (\cref{fig:rcwa_framework}e).

For the metalens-scintillator comparison, we evaluate both this full-wave RCWA model and a simplified design-agnostic model that applies the designed wavelength-dependent metalens phase; the latter is the same metasurface model used for the clinical detector simulations. The unity transmission and full-wave RCWA models give LSF FWHM values of 7.9~$\mu$m and 11.8~$\mu$m, respectively, compared with 22.8~$\mu$m experimentally (\cref{fig:metalens_simulation_experiment_comparison}a--c). Both models reproduce the substantial narrowing of the edge response. Including the wavelength-dependent nanopillar response increases the predicted LSF FWHM from \(7.9\) to \(11.8~\mu\mathrm{m}\), reducing the discrepancy from experiment from \(14.9\) to \(11.0~\mu\mathrm{m}\), or by approximately \(26\%\), although residual experimental broadening remains. Possible contributors include fabrication-induced variations in nanopillar diameter, height, and sidewall geometry, as well as aberrations or misalignment in the collection optics. In addition, the metalens is designed using the local periodic approximation, in which each nanopillar is assigned the RCWA response of an infinite periodic array of identical unit cells. Near the metalens edge, the phase wraps increasingly rapidly and adjacent nanopillars can differ substantially, violating this local-periodicity assumption. The resulting inter-element coupling and nonperiodic response are therefore not captured by the unit-cell RCWA model and may contribute to the difference between the simulated and measured resolution (and can also be observed in the darkening at the edges of the hemisphere). This is a limitation of the local-periodic design approach rather than of metalenses generally: advanced methods, including supercell-based optimization and full-wave inverse design, can explicitly account for interactions between dissimilar neighboring elements and thereby reduce errors associated with rapid phase variation.

Together, the bulk-scintillator agreement supports the depth-dependent scintillation propagation and finite-NA collection model, while the metalens-scintillator comparison shows that the framework predicts the experimentally observed substantial narrowing of the edge response and captures the trend introduced by the wavelength-dependent metasurface response. These results provide experimental support for applying the same numerical transport framework to the CT-relevant analysis, with the material properties, geometry, metasurface response, and detector parameters replaced by those of the clinical configuration. The resulting CdWO$_4$ detector predictions nevertheless remain projections for a CT-relevant architecture rather than a direct experimental validation of the complete clinical system.

\clearpage
\section{Experimental preparation and data post-processing methods}

\begin{minipage}{\textwidth}
    \centering
    \includegraphics[width=\textwidth]{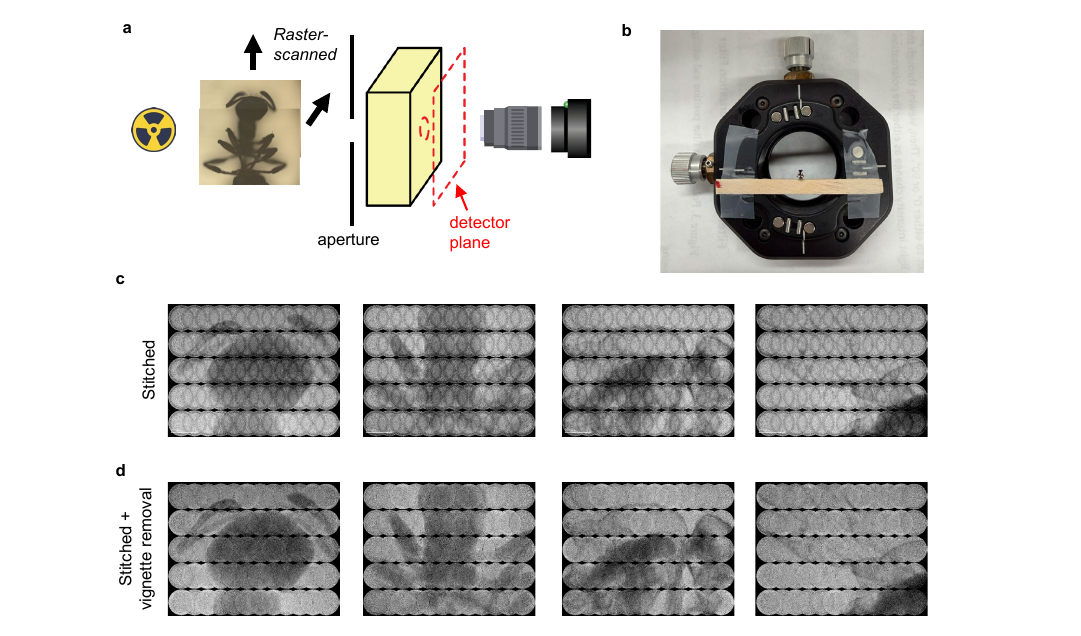}
    \captionof{figure}{\small
    \textbf{Raster-scanned specimen imaging and mosaic post-processing.}
    \textbf{a,} Experimental arrangement for raster-scanned imaging. The metalens--scintillator and specimen are translated using separate $x$--$y$ stages around the fixed aperture.
    \textbf{b,} Photograph of an iodine-stained ant mounted on a wooden support attached to the specimen translation stage.
    \textbf{c,} Mosaics obtained by placing the cropped circular metalens--scintillator images according to the commanded stage positions. Intensity reduction near the perimeter of each field produces visible vignetting and tile boundaries.
    \textbf{d,} Corresponding mosaics after radial vignetting correction, inter-tile brightness adjustment, and averaging of overlapping pixels. The fourth image shows an iodine-stained fruit-fly wing; this image is included here for completeness but is not shown in the main text because of its comparatively low contrast.}
    \label{fig:experimental_stitching}
    \vspace{0.3cm}
\end{minipage}

\paragraph{Specimen preparation.}
Adult ant specimens, identified as \textit{Tetramorium immigrans}, were field-collected from an outdoor environment in Cambridge, Massachusetts, and immediately preserved in ethanol. Adult fruit fly specimens, \textit{Drosophila melanogaster}, were provided by the Littleton Lab and similarly preserved in ethanol. The biological specimens used here were insects and did not involve vertebrate animal subjects. Ant and fruit fly specimens were transferred from ethanol into 5\% Lugol's iodine solution. To promote stain penetration, the immersed specimens were placed in a vacuum chamber and subjected to three evacuation--release cycles reaching approximately 28~inHg vacuum, followed by a 10-min vacuum treatment. The specimens remained immersed in Lugol's solution for five days, after which they were removed and air-dried for 48~h. Each specimen was mounted on a wooden support using adhesive, and the support was secured to an $x$--$y$ translation stage using scotch tape.

\paragraph{Raster-scanned imaging and stitching.}
The specimen and metalens--scintillator were mounted on separate, independently adjustable $x$--$y$ translation stages, with the aperture held fixed between them. The aperture and metalens were first aligned with the optical detection path and then held fixed while the specimen was raster-scanned through the approximately 250-$\mu$m-diameter metalens field of view (\cref{fig:experimental_stitching}a,b). Each mosaic contains 12 horizontal positions separated by 125~$\mu$m and five vertical positions separated by 250~$\mu$m, for a total of 60 images. Including the diameter of the circular field, these positions cover a nominal bounding region of $1.625~\mathrm{mm}\times1.25~\mathrm{mm}$.

For each position, the usable circular metalens field was cropped and placed according to the commanded stage displacement. Direct assembly of the cropped images produces visible circular boundaries because the collected intensity decreases toward the edge of each metalens field (\cref{fig:experimental_stitching}c). An empirical radial brightness correction was therefore applied to the vignetted outer diameter of each tile. Brightness was also adjusted for each tile to match the overall intensity of neighboring tiles within their overlapping regions. Pixels sampled by multiple tiles were combined by arithmetic averaging, reducing discontinuities at the tile boundaries and producing the processed stitched images shown in \cref{fig:experimental_stitching}d.

\clearpage
\section{Complete statistical DQE calculation pipeline}

\paragraph{Common simulation pipeline.}
The pixelated-scintillator and metalens--scintillator architectures are evaluated using the same architecture-agnostic Python functions and stochastic simulation pipeline. Shared physical inputs—including the X-ray spectrum, CdWO$_4$ thickness and attenuation coefficients, scintillation yield, detector quantum efficiency, and exposure timing—are held constant. More importantly, both architectures use the same computational methods and statistical procedure to calculate the MTF, NNPS, and DQE, including the same sequence of Poisson sampling, read-noise addition, ensemble averaging, Fourier analysis, and normalization. Only the architecture-specific PSFs, detector sampling, and electronic noise are substituted. This common implementation minimizes systematic differences arising from the numerical procedure and supports a controlled comparison of relative performance, though the absolute DQE values remain dependent on the stated physical assumptions.

\paragraph{Modulation transfer function.}
For each architecture, the response to a Dirac delta X-ray unit impulse is calculated using the scintillation wave-optics framework described in Supplementary Note~1. Let $\widetilde{h}_{i}(\mathbf{r})$ denote the unit-normalized intensity PSF for scintillation generated in depth plane $i$, where $\mathbf{r}$ is the transverse coordinate in the detector plane. Because emission from different scintillator depths is mutually incoherent, their intensities, rather than their complex fields, are summed. The depth-integrated optical PSF is therefore

\begin{equation}
h_{\mathrm{opt}}(\mathbf{r})
=
\sum_i
w_i\,\eta_i\,\widetilde{h}_{i}(\mathbf{r}),
\label{eq:depth_integrated_psf}
\end{equation}

where $w_i$ is the normalized Beer--Lambert energy-deposition weight and $\eta_i$ is the architecture-dependent optical throughput for that depth. For the metalens--scintillator, $\eta_i$ includes the depth-dependent fraction of isotropic emission collected directly or following reflection from the X-ray-incident surface. The energy-deposition-weighted collection efficiency is then 26.7\%.

The optical MTF is calculated from the magnitude of the Fourier transform of the depth-integrated PSF and normalized to unity at zero spatial frequency. 

Detector integration over the finite active area of each readout element is then included through the rectangular-aperture response,

\begin{equation}
\mathrm{MTF}_{\mathrm{sys}}(f_x,f_y)
=
\mathrm{MTF}_{\mathrm{opt}}(f_x,f_y)
\left|
\operatorname{sinc}(a_x f_x)
\operatorname{sinc}(a_y f_y)
\right|,
\label{eq:system_mtf}
\end{equation}

where $a_x$ and $a_y$ are the active detector dimensions and $\operatorname{sinc}(x)=\sin(\pi x)/(\pi x)$. The resulting two-dimensional MTF is azimuthally averaged to obtain the reported one-dimensional radial MTF. Spatial frequencies are first calculated in detector-plane coordinates and subsequently mapped to scanner isocenter using $f_{\mathrm{iso}}=M f_{\mathrm{det}}$, with $M=1.9$.

The depth-dependent geometric collection efficiency is incorporated by normalizing the sum of each PSF in the stack, such that convolution with the PSF performed in the NNPS calculation properly leads to less light collection in the metalens-scintillator flat-field images.

The depth-dependent geometric collection efficiency is retained in the PSF stack used for the noise calculation. Each unit-normalized metalens--scintillator PSF is rescaled such that its discrete sum equals the geometric collection fraction for that depth. The resulting energy-deposition-weighted collection efficiency is 26.7\%. For the conventional architecture, each depth-dependent PSF is represented by a unit-sum spatial delta function. These PSF normalizations propagate the architecture-dependent light collection into the mean detected signal and therefore into the subsequent NNPS and DQE calculations.

\paragraph{Stochastic flat-field simulation.}
The NNPS is calculated from ensembles of uniformly irradiated flat-field images using the standard Fourier-domain approach for X-ray detector noise analysis~\cite{nhsbsp2025ffdm}. The incident fluence rate, $\dot{\Phi}_0$, is converted to the incident fluence per projection according to

\begin{equation}
\overline{q}
=
\dot{\Phi}_0
\frac{0.33~\mathrm{s}}{1000},
\label{eq:fluence_per_projection}
\end{equation}

where $\overline{q}$ has units of photons\,mm$^{-2}$ per projection. Fluence rates from $10^3$ to $10^9$ photons\,mm$^{-2}$\,s$^{-1}$ are evaluated, spanning highly attenuated anatomical paths to open-beam clinical CT conditions as described in Supplementary Note~2. For the conventional detector, the mean number of incident X-rays within each sampling element is additionally multiplied by the geometric fill factor, $\eta_{\mathrm{fill}}=0.806$. The metalens--scintillator is assumed to have no inactive separator area.

For each source-energy bin $k$ and scintillator depth interval $i$, the expected number of locally absorbed X-ray interactions within a sampling element is

\begin{equation}
\Lambda_{i,k}
=
\overline{q}\,A_{\mathrm{samp}}\,s_k
\left[
\exp\!\left(-\mu_k z_i\right)
-
\exp\!\left(-\mu_k(z_i+\Delta z)\right)
\right]
\frac{\mu_{\mathrm{en},k}}{\mu_k},
\label{eq:expected_xray_interactions}
\end{equation}

where $A_{\mathrm{samp}}$ is the detector-plane sampling area, $s_k$ is the normalized spectral fluence in energy bin $k$, and $\mu_k$ and $\mu_{\mathrm{en},k}$ are the corresponding linear attenuation and energy-absorption coefficients. Independent Poisson draws,

\begin{equation}
N_{i,k}(\mathbf{r})
\sim
\operatorname{Poisson}\!\left(\Lambda_{i,k}\right),
\label{eq:xray_poisson}
\end{equation}

are performed at every spatial location, energy, and depth to represent fluctuations in X-ray arrival and absorption.

The expected number of scintillation photons generated in depth plane $i$ is then

\begin{equation}
L_i(\mathbf{r})
=
\sum_k
Y E_k N_{i,k}(\mathbf{r}),
\label{eq:scintillation_generation}
\end{equation}

where $Y=28$ photons\,keV$^{-1}$ is the assumed CdWO$_4$ scintillation yield~\cite{michail2020luminescence} and $E_k$ is the X-ray energy. A fixed mean light yield is used; no additional empirical Swank factor is introduced, re-iterating the importance that the architectures are evaluated using the same computational pipeline.

Optical transport is applied independently at every depth by convolving the generated photon distribution with the corresponding PSF:

\begin{equation}
\Lambda_{\gamma}(\mathbf{r})
=
\sum_i
\left[
h_i * L_i
\right](\mathbf{r}),
\label{eq:flatfield_optical_transport}
\end{equation}

where $*$ denotes two-dimensional convolution. For the metalens--scintillator, the sum of $h_i$ equals the geometric collection fraction at depth $i$, thereby accounting simultaneously for spatial redistribution and collection loss. For the conventional detector, $h_i$ has unit sum regardless of scintillator depth. The resulting image, $\Lambda_{\gamma}(\mathbf{r})$, is the expected optical-photon distribution at the detector.

The discrete number of optical photons incident on each detector element is sampled as

\begin{equation}
N_{\gamma}(\mathbf{r})
\sim
\operatorname{Poisson}\!\left[
\Lambda_{\gamma}(\mathbf{r})
\right].
\label{eq:optical_poisson}
\end{equation}

Applying this Poisson draw after convolution is equivalent to independently distributing Poisson-distributed optical photons among the detector elements according to the depth-dependent PSFs, while avoiding explicit tracking of every scintillation photon. Detector sensor photoelectron generation is subsequently modeled as

\begin{equation}
N_{\mathrm{e}}(\mathbf{r})
\sim
\operatorname{Binomial}
\left[
N_{\gamma}(\mathbf{r}),
\mathrm{QE}
\right],
\label{eq:qe_binomial}
\end{equation}

with $\mathrm{QE}=0.94$. Finally, zero-mean Gaussian electronic read noise is added:

\begin{equation}
I_m(\mathbf{r})
=
N_{\mathrm{e}}(\mathbf{r})
+
\epsilon_m(\mathbf{r}),
\qquad
\epsilon_m
\sim
\mathcal{N}\!\left(0,\sigma_{\mathrm{read}}^2\right),
\label{eq:read_noise}
\end{equation}

where $\sigma_{\mathrm{read}}$ is 1.87$\times10^3$ electrons RMS for the conventional detector and 7.2 electrons RMS for the metalens--scintillator detector, as described in Supplementary Note~2.

\paragraph{Noise power spectrum.}
For each architecture and fluence rate, $M_{\mathrm{r}}=50$ statistically independent flat-field realizations are generated. The spatial mean of each realization is subtracted to remove its zero-frequency component:

\begin{equation}
\Delta I_m(\mathbf{r})
=
I_m(\mathbf{r})
-
\left\langle I_m(\mathbf{r})\right\rangle_{\mathbf{r}}.
\label{eq:flatfield_fluctuation}
\end{equation}

For an image containing $N_x\times N_y$ samples with detector-plane sampling intervals $\Delta x$ and $\Delta y$, the two-dimensional NPS is calculated as

\begin{equation}
\mathrm{NPS}(f_x,f_y)
=
\frac{\Delta x\,\Delta y}{N_xN_y}
\frac{1}{M_{\mathrm{r}}}
\sum_{m=1}^{M_{\mathrm{r}}}
\left|
\mathcal{F}_{\mathrm{d}}
\left\{
\Delta I_m(\mathbf{r})
\right\}
\right|^2,
\label{eq:nps}
\end{equation}

where $\mathcal{F}_{\mathrm{d}}$ denotes the unnormalized discrete Fourier transform. The ensemble-averaged signal is

\begin{equation}
\overline{I}
=
\frac{1}{M_{\mathrm{r}}}
\sum_{m=1}^{M_{\mathrm{r}}}
\left\langle I_m(\mathbf{r})\right\rangle_{\mathbf{r}},
\label{eq:mean_flatfield_signal}
\end{equation}

and the normalized noise power spectrum is

\begin{equation}
\mathrm{NNPS}(f_x,f_y)
=
\frac{\mathrm{NPS}(f_x,f_y)}
{\overline{I}^{\,2}}.
\label{eq:nnps}
\end{equation}

The NNPS therefore has units of mm$^2$. The two-dimensional spectrum is azimuthally averaged to obtain the reported radial NNPS, using the same frequency bins as the corresponding MTF.

\paragraph{Detective quantum efficiency.}
The frequency-dependent DQE is calculated from the system MTF, NNPS, and incident X-ray fluence per projection using the standard expression~\cite{tsai2011mutual,garcia2011study}

\begin{equation}
\mathrm{DQE}(f)
=
\frac{
\mathrm{MTF}_{\mathrm{sys}}^2(f)
}{
\overline{q}\,\mathrm{NNPS}(f)
}.
\label{eq:dqe}
\end{equation}

Here, $\overline{q}$ is the fluence incident on the detector before application of the conventional-detector fill factor or any architecture-dependent optical losses. Consequently, reductions in active area, optical collection, detector quantum efficiency, and electronic noise performance are retained in the output NNPS and directly reduce the calculated DQE.

At high fluence, DQE is primarily limited by X-ray absorption, optical transport, and detector sampling. At lower fluence, the signal-dependent quantum noise decreases while the electronic read-noise variance remains constant, causing the DQE to decrease. The DQE is calculated in detector-plane spatial-frequency coordinates and then reported at scanner isocenter using $f_{\mathrm{iso}}=M f_{\mathrm{det}}$, with $M=1.9$. This coordinate transformation changes the reported spatial-frequency axis but not the dimensionless DQE value. Results are reported only within the Nyquist bandwidth of the corresponding detector architecture.

\clearpage
\section{Task-weighted detectability score and dose-reduction analysis}

\begin{minipage}{\textwidth}
    \centering
    \includegraphics[width=0.7\textwidth]{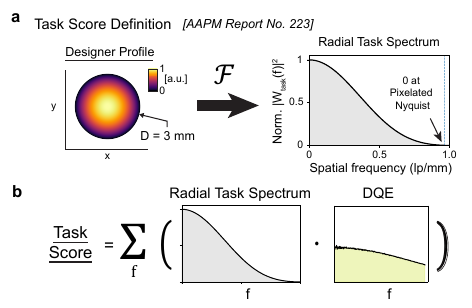}
    \captionof{figure}{\small
    \textbf{Task-weighted detectability-score calculation.}
    \textbf{a,} A radially symmetric 3-mm-diameter designer profile and its corresponding task spectrum. The task was selected such that its dominant spatial-frequency content lies below the Nyquist frequency of the pixelated scintillator.
    \textbf{b,} The task-weighted score is obtained by integrating the product of the radial task power spectrum, DQE, and polar-frequency weighting factor $2\pi f$.}
    \label{fig:task_weighted_detectability}
    \vspace{0.3cm}
\end{minipage}

\paragraph{Task definition.}
To quantify the potential dose reduction offered by the metalens--scintillator architecture, we evaluated a task-weighted detectability score at each simulated photon fluence rate. Following the task-function framework described in AAPM Report No.~233, \textit{Performance Evaluation of Computed Tomography Systems}~\cite{samei2019performance}, we represented the target feature using a radially symmetric designer profile (\cref{fig:task_weighted_detectability}a):
\begin{equation}
c(r)=
\begin{cases}
C\left[1-\left(\dfrac{2r}{D}\right)^2\right]^n, & 0\leq r\leq D/2,\\[6pt]
0, & r>D/2,
\end{cases}
\label{eq:designer_profile}
\end{equation}
where $D$ is the physical diameter of the target, $C$ is its peak contrast, and $n$ controls the edge roll-off. We used $n=1$, as recommended in AAPM Report No.~233, and set $C=1$ because only relative scores between detector architectures were compared.

The task function, $W_{\mathrm{task}}$, was defined as the two-dimensional Fourier transform of $c(r)$. For the $n=1$ profile used here, its radial form is
\begin{equation}
W_{\mathrm{task}}(f)
=
C\pi D^2
\frac{J_2(\pi Df)}{(\pi Df)^2},
\label{eq:designer_task_spectrum}
\end{equation}
where $J_2$ is the second-order Bessel function of the first kind and $W_{\mathrm{task}}(0)=C\pi D^2/8$.

\paragraph{Task-weighted DQE score.}
We defined the task-weighted DQE score as the integral of the task power spectrum weighted by the system DQE over the two-dimensional spatial-frequency domain:
\begin{equation}
S_{\mathrm{task}}
=
\int_{-\infty}^{\infty}
\int_{-\infty}^{\infty}
\left|W_{\mathrm{task}}(f_x,f_y)\right|^2
\operatorname{DQE}(f_x,f_y)
\,df_x\,df_y,
\label{eq:task_score_2d}
\end{equation}
where $f_x$ and $f_y$ are the Cartesian spatial frequencies. Because both the designer profile and the simulated DQE were radially symmetric, this two-dimensional integral reduced to
\begin{equation}
S_{\mathrm{task}}
=
\int_{0}^{f_{\mathrm{N}}}
\left|W_{\mathrm{task}}(f)\right|^2
\operatorname{DQE}(f)\,2\pi f\,df,
\label{eq:task_score_radial}
\end{equation}
where $f_{\mathrm{N}}$ is the applicable Nyquist frequency. The factor $2\pi f$ accounts for the polar area element in the two-dimensional spatial-frequency domain. The resulting score therefore weights detector performance most strongly at spatial frequencies that contribute to the specified imaging task (\cref{fig:task_weighted_detectability}b).

\paragraph{Task selection.}
To compare dose performance without imposing an aliasing penalty on the reference detector, we selected a task with a diameter of $D=3~\mathrm{mm}$ whose dominant spatial-frequency content lies below the Nyquist frequency of the pixelated scintillator. The 0.98-mm detector pitch corresponds to a detector-plane Nyquist frequency of approximately $0.51~\mathrm{lp\,mm^{-1}}$, or $0.97~\mathrm{lp\,mm^{-1}}$ at isocenter after applying the clinical geometric magnification of $M=1.9$. The selected task can therefore be fully resolved by both architectures and provides a common basis for evaluating their dose-dependent performance.

A brief analogy to explain the above: if we use a task that extends beyond the Nyquist frequency of the pixelated scintillator, ``It’s like asking two people to read an eye chart in the dark, making the letters too small for one person to resolve no matter how bright the room is, and then claiming the other person has million-times better night vision.''

\paragraph{Dose-reduction evaluation.}
We calculated $S_{\mathrm{task}}$ across the simulated range of incident photon fluence rates to generate dose-response curves for both architectures. At $10^5~\mathrm{photons\,mm^{-2}\,s^{-1}}$, representative of a strongly attenuated CT projection, the pixelated scintillator produced the reference task score. The metalens--scintillator achieved the same score at approximately $2\times10^4~\mathrm{photons\,mm^{-2}\,s^{-1}}$, as determined by log--log interpolation between the simulated exposure levels. For a fixed acquisition time, beam spectrum, and imaging geometry, this fivefold reduction in incident fluence rate corresponds to a fivefold reduction in incident photon fluence, and hence radiation dose, while preserving the modeled detectability of the 3-mm task. This improvement was obtained despite a geometric light-collection efficiency of only 26.7\%, indicating further scope for optimization through more efficient optical architectures or computational reconstruction.

This dose-reduction estimate deliberately considers a task whose spatial-frequency content is effectively confined below the Nyquist frequency of the reference pixelated scintillator. It therefore isolates the dose-efficiency advantage of the metalens--scintillator for a feature that both architectures can resolve and does not account for the substantially broader spatial-frequency bandwidth of the metalens--scintillator. For tasks containing finer spatial features beyond the pixelated-detector Nyquist limit, as shown in Fig.~4c and d, the metalens--scintillator provides an additional resolution advantage that is not captured by the fivefold dose-reduction estimate reported here.

\clearpage
\section{Statistical simulation model of trabecular bone, lung tissue, and breast microcalcifications}

\paragraph{Simulation overview.}
We extended the statistical detector framework described in Supplementary Note~9 from uniform flat-field and point-response inputs to spatially varying anatomical projections. The anatomical samples are represented by two-dimensional material path-length maps. These maps were used to calculate the energy-dependent X-ray transmission at each spatial position, after which the transmitted photons were propagated through the same depth-resolved scintillation and detector model used for the DQE calculations. The depth-dependent metalens point-spread functions (PSFs) were obtained from the angular-spectrum framework described in Supplementary Note~1. The resulting images represent individual X-ray projections rather than reconstructed CT volumes.

\paragraph{Anatomical input models.}
Three anatomical models were constructed to represent structures spanning different materials, morphologies, and spatial scales.

\textit{Trabecular bone.}
The trabecular-bone model was constructed from the LHDL human trabecular-bone micro-CT dataset~\cite{Iori2023}. The source volume had an isotropic voxel size of $19.5~\mu\mathrm{m}$ and a physical extent of $3.90~\mathrm{mm}$ along each dimension. Voxels with grayscale values greater than 63 were classified as mineralized bone, and the resulting binary mask was summed along the projection direction to produce a hydroxyapatite path-length map. The bone was modeled as hydroxyapatite with a density of $3.16~\mathrm{g\,cm^{-3}}$ embedded in water with a density of $1.00~\mathrm{g\,cm^{-3}}$. Hydroxyapatite was assumed to displace water, such that the combined water and hydroxyapatite path lengths at each position equaled the total sample thickness. The resulting path-length maps were resampled to the $4.13~\mathrm{mm}\times4.13~\mathrm{mm}$ simulation field of view from $3.9~\mathrm{mm}\times3.9~\mathrm{mm}$ to match the metalens-scintillator field of view. 

\textit{Lung tissue.}
The lung model was constructed from the HiP-CT human left-lung VOI-05 dataset~\cite{https://doi.org/10.15151/esrf-dc-572229315}, which has an isotropic voxel size of $10~\mu\mathrm{m}$. A $4.13~\mathrm{mm}\times4.13~\mathrm{mm}$ region was selected from slices 01738--01864. Voxels with grayscale values greater than or equal to 5802 were classified as soft tissue, whereas the remaining voxels were classified as air. The binary tissue mask was integrated along the projection direction to generate a soft-tissue path-length map. The projected thickness was scaled by a factor of four to obtain a maximum equivalent sample thickness of $5.08~\mathrm{mm}$. Soft tissue was assigned a density of $1.06~\mathrm{g\,cm^{-3}}$, while air was treated as having negligible attenuation. This simplified binary model was selected to preserve the fine septal and vascular morphology of the source volume while providing a well-defined soft-tissue--air contrast.

\textit{Breast microcalcifications.}
The breast-microcalcification model was constructed from a selected region of mammogram 53582791 in the INbreast dataset~\cite{moreira2012inbreast}. A $4.13~\mathrm{mm}\times4.13~\mathrm{mm}$ region containing a microcalcification cluster was resampled to a $13.5~\mu\mathrm{m}$ pixel pitch. Because a mammogram is a two-dimensional image, it does not directly provide material thickness along the X-ray propagation direction. We therefore used the image to construct separate projected-thickness maps for the calcifications and surrounding breast tissue.

To construct the calcification thickness map, pixels with grayscale values greater than or equal to 168 were classified as calcifications. Because their out-of-plane thickness was unknown, we assumed that the maximum calcification thickness was equal to the maximum calcification width measured in the mammogram. This width, $538.3~\mu\mathrm{m}$, was therefore used as the maximum hydroxyapatite thickness. Within the calcification mask, grayscale values were mapped linearly from zero thickness at the segmentation threshold of 168 to a thickness of $538.3~\mu\mathrm{m}$ at the maximum grayscale value of 255.

To construct the breast-tissue thickness map, the calcification pixels were first replaced with values estimated from the surrounding background. This prevented the calcification signal from being included in both material maps. The resulting background grayscale values were then mapped linearly to breast-tissue thicknesses between 0.8 and $1.2~\mathrm{mm}$. Breast tissue and hydroxyapatite were assigned densities of $1.02~\mathrm{g\,cm^{-3}}$ and $3.16~\mathrm{g\,cm^{-3}}$, respectively.

This procedure provides a projected-thickness model for comparing the two detector architectures. It is not intended to reproduce the true three-dimensional shapes or absolute thicknesses of the calcifications.

\paragraph{Polychromatic object transmission.}
For each anatomical model, the energy-dependent transmitted photon fluence was calculated using the Beer--Lambert law:
\begin{equation}
\Phi_{\mathrm{trans}}(E,x,y)
=
\Phi_0(E)
\exp\left[
-\sum_m \mu_m(E)t_m(x,y)
\right],
\label{eq:anatomical_transmission}
\end{equation}
where $\Phi_0(E)$ is the incident spectral photon fluence, $\mu_m(E)$ is the linear attenuation coefficient of material $m$, and $t_m(x,y)$ is its projected path length. Material attenuation coefficients were obtained from the NIST database and were interpolated onto the energy bins of the filtered 120-kVp spectrum as described in Supplementary Note~2.

The transmitted spectrum at each spatial position was used as the input to the 2.3-mm-thick CdWO$_4$ scintillator model. The scintillator was discretized into 101 depth planes. For energy bin $E$ and depth interval extending from $z_i$ to $z_{i+1}$, the expected number of X-ray interactions was
\begin{equation}
\lambda_{i,E}(x,y)
=
A\,\Phi_{\mathrm{trans}}(E,x,y)
\left[
e^{-\mu_{\mathrm{s}}(E)z_i}
-
e^{-\mu_{\mathrm{s}}(E)z_{i+1}}
\right],
\label{eq:anatomical_depth_interactions}
\end{equation}
where $A$ is the area of one simulation-grid element and $\mu_{\mathrm{s}}(E)$ is the linear attenuation coefficient of CdWO$_4$. Independent Poisson samples were drawn from $\lambda_{i,E}(x,y)$ to model the stochastic number of interactions in each energy and depth bin. The locally deposited energy was calculated by multiplying each interaction by
\begin{equation}
E\frac{\mu_{\mathrm{en,s}}(E)}{\mu_{\mathrm{s}}(E)},
\end{equation}
where $\mu_{\mathrm{en,s}}(E)$ is the CdWO$_4$ linear energy-absorption coefficient. The deposited energy was converted to scintillation light using a yield of 28 optical photons per keV, as outlined in Supplementary Note~2.

\paragraph{Depth-resolved optical image formation.}
Using the framework developed in Supplementary Note~9, the optical source distribution generated at each scintillator depth was multiplied by the corresponding lateral-position and depth-dependent metalens light collection fraction, and convolved with the depth-specific PSF calculated in Supplementary Note~1. The expected optical-photon distribution at the detector was therefore
\begin{equation}
\Lambda_{\gamma}(x,y)
=
\sum_i
h_i(x,y)
*
\left[
\eta_i(x,y)L_i(x,y)
\right],
\label{eq:anatomical_metalens_image}
\end{equation}
where $L_i(x,y)$ is the stochastic scintillation source distribution in depth plane $i$, $\eta_i(x,y)$ is the direct-plus-reflected geometric collection fraction, $h_i(x,y)$ is the corresponding unit-normalized intensity PSF, and $*$ denotes two-dimensional convolution. Because scintillation generated at different depths is mutually incoherent, the convolved intensity distributions were summed.

The resulting optical distribution was integrated onto a detector with a $13.5~\mu\mathrm{m}$ pixel pitch. As explained in Supplementary Note~9, an additional Poisson draw was applied to the expected number of optical photons incident on each detector pixel. Photoelectron conversion was modeled by binomial sampling with a quantum efficiency of 0.94, followed by additive Gaussian read noise with a standard deviation of 7.2 electrons RMS, following the parameters from Supplementary Note~2. The simulations used the open-beam incident photon fluence rate of $10^9~\mathrm{photons\,mm^{-2}\,s^{-1}}$. A simulated flat-field image was generated using the corresponding uniform background material and propagated through the same position-dependent collection and PSF model. The object image was divided by this flat-field response to correct the spatial collection efficiency of the metalens-scintillator.

\paragraph{Detector-image comparison.}
For each anatomical model, three images were generated for comparison in Fig.~5. The ground-truth image was defined as the expected transmitted and absorbed X-ray signal before optical blurring or detector sampling. The conventional pixelated-scintillator image was generated by integrating this expected signal over the active areas of a $4\times4$ array with a 0.98-mm pitch and 100-$\mu$m-wide inactive separators. The metalens--scintillator image included the complete depth-resolved PSF, optical-collection, photon-counting, quantum-efficiency, and read-noise model described above.

For visualization, the grayscale limits of each image were set independently to its first and 99th intensity percentiles. This normalization was applied only to the displayed images and does not alter the underlying simulated detector signals. 

\paragraph{Scope and numerical validation.}
This reduced-order statistical framework retains the principal processes governing polychromatic object attenuation, depth-dependent scintillation generation, optical collection, spatial resolution, quantum noise, and electronic read noise without explicitly tracking individual particle trajectories and optical rays. Once the depth-dependent PSFs have been calculated, this approach generates a simulated projection in seconds, compared with the hours required for the corresponding particle- and ray-tracing simulations. Supplementary Note~11 compares the predictions of this framework with a more complete Geant4--Zemax model, in which X-ray interactions, scintillation-photon generation, and subsequent optical propagation are simulated at the particle and ray levels.

\clearpage
\section{Geant4 validation of statistical simulation}

\begin{minipage}{\textwidth}
    \centering
    \includegraphics[width=0.9\textwidth]{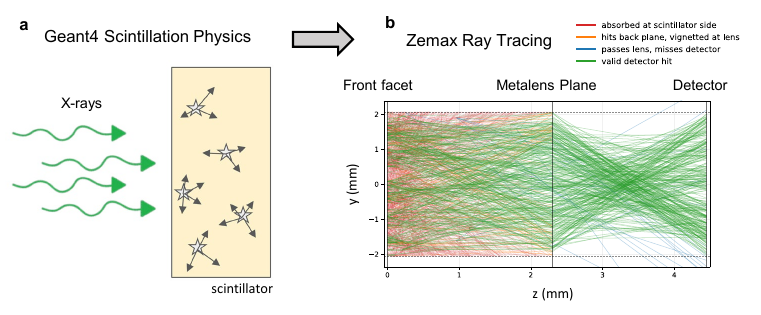}
    \captionof{figure}{\small
    \textbf{Combined Geant4--Zemax simulation framework.}
    \textbf{a,} Geant4 simulation of X-ray interactions and scintillation-photon generation within the CdWO$_4$ scintillator. The position and emission direction of each generated optical photon are recorded for subsequent optical propagation.
    \textbf{b,} Representative trajectories of sampled optical photons propagated through the scintillator, metalens, and detector using Zemax ray tracing. Photons traveling toward the front facet are specularly reflected, whereas photons intersecting a lateral scintillator boundary are treated as lost. The colors distinguish photons lost at the scintillator sides, photons vignetted at the metalens aperture, photons that pass through the metalens but miss the detector, and photons that produce valid detector hits.}
    \label{fig:geant4_zemax_pipeline}
    \vspace{0.3cm}
\end{minipage}

Geant4 is a Monte Carlo simulation toolkit that models scintillation by calculating the energy deposited through interactions between high-energy particles and matter~\cite{agostinelli2003geant4}. Although Geant4 provides a powerful framework for modeling scintillation processes at the particle-interaction level, it does not intrinsically account for the nanophotonic effects introduced by a focusing metalens. We therefore combine Geant4 and Zemax simulations to model collected images of different medical samples: Geant4 is used to model X-ray transport through the samples and the subsequent generation of optical photons within the scintillator, while Zemax is used to model the propagation of these optical photons through the metalens system to the detector.

\paragraph{Preparation of simulation tissue samples.}
Three different medical samples (trabecular bone, lung tissue, and breast microcalcifications) are modeled as 3D voxelized objects. Each voxel is assigned the corresponding material defined in the previous section. The material composition and density are specified according to the chemical formula and mass density of each constituent material. All samples have the same physical area of $4.13~\mathrm{mm} \times 4.13~\mathrm{mm}$. Each sample is centered at the origin of the Geant4 simulation geometry. 

\paragraph{Geant4 simulation details.}
A planar X-ray source with dimensions of $4.13~\mathrm{mm} \times 4.13~\mathrm{mm}$ is used to generate X-ray images of medical samples. The experimentally relevant CT X-ray spectrum is implemented in the simulation, as outlined in Supplementary Note~2. The material parameters governing scintillation in CdWO$_4$, including its scintillation yield and density are set according to the values described in Supplementary Note~2. The CdWO$_4$ scintillator has dimensions of $4.13~\mathrm{mm} \times 4.13~\mathrm{mm} \times 2.3~\mathrm{mm}$. Electromagnetic interactions and optical processes are modeled using \texttt{G4EmStandardPhysics\_option4} and \texttt{G4OpticalPhysics}, respectively.

For each medical sample, the spatial position, wavelength, and emission direction of every generated optical photon are recorded at the point of generation within the scintillator. These photon data are subsequently used as inputs to the Zemax simulation to model optical propagation through the metalens and to reconstruct the corresponding X-ray image at the detector. A simulated flat-field image was calculated using the corresponding uniform background material and propagated through the same Geant4 and Zemax framework. The object image was divided by this flat-field response to correct for the spatial collection variation of the metalens-scintillator.
 
\paragraph{Zemax photon propagation.}
For each optical photon generated in Geant4, the generation position $(x,y,z)$, momentum components $(p_x,p_y,p_z)$, and wavelength were stored. The photon positions were transformed into the Zemax coordinate system. Photons initially traveling toward the metalens were propagated directly. Photons traveling toward the X-ray entrance surface were specularly reflected by the ideal front reflector before being propagated toward the metalens, whereas photons that intersected the lateral scintillator boundaries were treated as lost.

\begin{minipage}{\textwidth}
    \centering
    \includegraphics[width=0.85\textwidth]{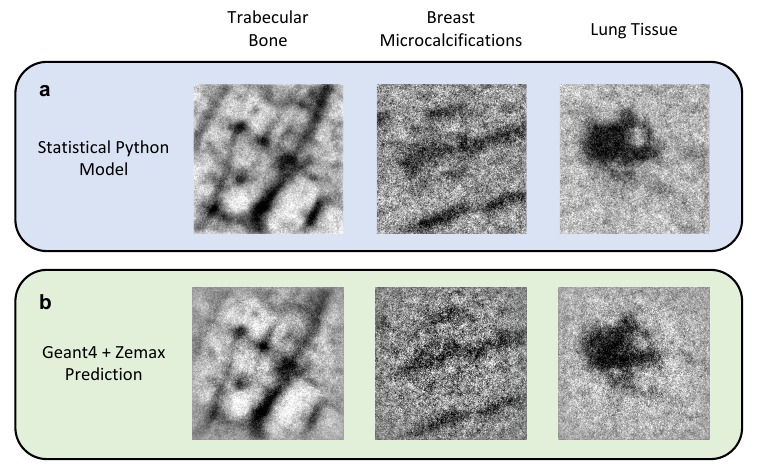}
    \captionof{figure}{\small
    \textbf{Validation of the statistical image-formation model against combined Geant4--Zemax simulations.}
    Representative simulated metalens-scintillator images of trabecular bone, breast microcalcifications, and lung tissue are shown.
    \textbf{a,} Images generated using the statistical Python model, which accounts for depth-dependent X-ray energy deposition, optical collection, spatial blurring, and detector noise.
    \textbf{b,} Corresponding images generated using explicit Geant4 modeling of X-ray interactions and scintillation-photon generation followed by Zemax propagation of the optical photons through the metalens system.
    The qualitative agreement in sample morphology, spatial resolution, contrast, and noise characteristics supports the use of the computationally efficient statistical model for the scintillation framework presented in Supplementary Notes 1 and 9, and image simulations presented in the main text.}
    \label{fig:geant4_statistical_validation}
    \vspace{0.3cm}
\end{minipage}

The remaining photons were imported into Zemax OpticStudio through the ZOS-API and traced as unpolarized rays. The sequential optical model included propagation through the higher-index CdWO$_4$ scintillator, followed by a planar grid-phase surface at the rear scintillator interface that imposed the ideal spherical phase profile of the metalens over a $4.13~\mathrm{mm}\times4.13~\mathrm{mm}$ clear aperture. After the phase surface, rays were propagated to the detector plane. Rays that fell outside the metalens aperture or did not reach the detector were not recorded at the detector.

Given the incoherent nature of scintillation emission, the detected rays at the detector were summed incoherently. The detector binned into a $306\times306$ pixel array spanning $4.13~\mathrm{mm}\times4.13~\mathrm{mm}$, corresponding to a pixel pitch of approximately $13.5~\mu\mathrm{m}$. Each detected optical photon contributed one count to the corresponding detector pixel, and the final image was formed by incoherently summing all detected photons.


The same Geant4--Zemax pipeline was applied to both the medical sample and a corresponding uniform flat-field sample. The flat-field-corrected transmission image was calculated as $T(x,y)=I_{\mathrm{sample}}(x,y)/I_{\mathrm{flat}}(x,y)$, where $I_{\mathrm{sample}}$ and $I_{\mathrm{flat}}$ are the sample and flat-field detector images, respectively. This correction removes spatial variations in optical collection and vignetting introduced by the finite metalens aperture. The qualitative comparison of the two approaches is show in~\cref{fig:geant4_statistical_validation}.

\clearpage
\bibliographystyle{ieeetr}
\bibliography{bibliography}